\documentclass[aps,twocolumn,superscriptaddress,notitlepage,nofootinbib]{revtex4-1}

\usepackage{amsmath,bm}
\usepackage{graphicx}
\usepackage{epstopdf}
\usepackage{subfigure}
\usepackage{epsfig}
\usepackage{amsmath,amssymb,amsfonts}
\usepackage{color}
\usepackage[utf8]{inputenc}
\usepackage{verbatim}
\usepackage{array}
\usepackage{hyperref}
\usepackage{ulem}
\usepackage{multirow}

\begin{document}

\title{$J/\psi$ production within a jet in high-energy proton-proton and nucleus-nucleus collisions}

\date{\today  \hspace{1ex}}

\author{Shan-Liang Zhang}
\affiliation{Department of Physics, Hubei University, Wuhan 430062, China}
\affiliation{Key Laboratory of Atomic and Subatomic Structure and Quantum Control (MOE), Guangdong Basic Research Center of Excellence for Structure and Fundamental Interactions of Matter, Institute of Quantum Matter, South China Normal University, Guangzhou 510006, China}
\affiliation{Guangdong-Hong Kong Joint Laboratory of Quantum Matter, Guangdong Provincial Key Laboratory of Nuclear Science, Southern Nuclear Science Computing Center, South China Normal University, Guangzhou 510006, China}

\author{Hongxi Xing}
\email{hxing@m.scnu.edu.cn}
\affiliation{Key Laboratory of Atomic and Subatomic Structure and Quantum Control (MOE), Guangdong Basic Research Center of Excellence for Structure and Fundamental Interactions of Matter, Institute of Quantum Matter, South China Normal University, Guangzhou 510006, China}
\affiliation{Guangdong-Hong Kong Joint Laboratory of Quantum Matter, Guangdong Provincial Key Laboratory of Nuclear Science, Southern Nuclear Science Computing Center, South China Normal University, Guangzhou 510006, China}
\affiliation{Southern Center for Nuclear-Science Theory (SCNT), Institute of Modern Physics, Chinese Academy of Sciences, Huizhou 516000, China}

\begin{abstract}
Within the framework of leading power factorization formalism of nonrelativistic quantum chromodynamics, we calculate the jet fragmentation function for $J/\psi$ production in proton-proton (pp) collisions ranging from $\sqrt{s}=500$ GeV to $13$ TeV. The reasonable agreements between theory and experimental data indicate that $J/\psi$ production within a jet is mainly dominated by gluon fragmentation. Such a mechanism can be further tested by the predicted jet transverse momentum and radius dependence of jet fragmentation function. Based on the satisfying description of pp baseline, we carry out the first theoretical investigation on medium modification on $J/\psi$ production within jet in heavy-ion collisions at the Large Hadron Collider, using a linear Boltzmann transport model combined with hydrodynamics for the simulation of jet-medium interaction. The consistency with the experimental measurement on nuclear modification factor $R_{\text{AA}}$ by CMS collaboration reveals that the gluon jet quenching is the driving force for the suppression of $J/\psi$ production in jet. Furthermore, we make predictions for the dependence of $R_{\text{AA}}$ on the jet transverse momentum and jet radius $R$, which can be tested in future measurements to further constrain the flavor dependence of jet quenching.

\end{abstract}


\maketitle

\section{Introduction}
\label{sec:Intro}
The underlying mechanism for producing $J/\psi$ mesons as bound states of charm quark pairs ($c\bar{c}$) has been under intensive experimental and theoretical studies in both hadronic and heavy-ion collisions \cite{Andronic:2015wma,Chapon:2020heu,Lansberg:2019adr}.
The production of heavy quarkonium in high-energy particle collisions not only provide valuable information on non-perturbative QCD physics of hadronization, but also linked to longstanding challenges
in our understanding of the properties of QGP, a color-deconfined QCD medium, in relativistic heavy-ion collisions~\cite{Matsui:1986dk}.
Furthermore, it is proposed that the production yields of quarkonium at electron ion collisions are critical for studying the production mechanism and also can be used as an ideal tool to probe the gluon structure of nucleon and nucleus \cite{Flore:2020jau,DAlesio:2021yws,Boer:2021ehu,Chen:2023hvu,Qiu:2020xum}.

The non-relativistic QCD (NRQCD) factorization formalism \cite{Bodwin:1994jh} is by far the most phenomenologically successful framework to describe heavy quarkonium production in hadronic collisions, although there are still challenges to simultaneously describe their cross sections and polarisations \cite{Dai:2017cjq}. In recent years, quarkonium production within jets has been proposed as a probe of heavy-quark hadronization mechanisms \cite{Bain:2016clc,Baumgart:2014upa}, and was expected to give a new insight into this puzzle~\cite{Kang:2017yde}. Recently, the Large Hadron Collider (LHC) and Relativistic Heavy Ion Collider (RHIC) measured $J/\psi$ mesons produced in jet in proton-proton (pp) collisions. In particular, the transverse momentum fraction of jet, $z_{J/\psi}=p_\text{T}^{J/\psi}/p_\text{T}^{\text{jet}}$, carried by the $J/\psi$ have been measured in pp collisions by LHCb~\cite{LHCb:2017llq}, CMS~\cite{CMS:2021puf,CMS:2019ebt} and STAR~\cite{Yang:2021smr}(preliminary). These measurements indicate significant deviations from default Monte Carlo PYTHIA simulations~\cite{LHCb:2017llq,CMS:2021puf,Yang:2021smr}, which use leading order NRQCD complemented with subsequent parton shower~\cite{Sjostrand:2014zea}. Theoretical predictions, provided in~\cite{Bain:2016clc} by using the fragmenting jet functions (FJF) and the Gluon Fragmentation Improved PYTHIA (GFIP)\cite{Bain:2017wvk,Bain:2016clc}, which correspond to a modified parton shower approach where the quarkonium fragmentation is only implemented at the end of the shower, provide well description of the LHCb and CMS data of $J/\psi$ production in jets\cite{CMS:2019ebt,LHCb:2017llq}.

In heavy-ion collision, heavy quarkonia production at low transverse momentum has long been regarded as a signature of color deconfinement and a thermometer
of QGP \cite{Matsui:1986dk}. The interpretation of the experimental data for $J/\psi$ suppression observed both at RHIC~\cite{PHENIX:2006gsi,PHENIX:2011img} and the LHC ~\cite{NA50:2004sgj,ALICE:2012jsl,ALICE:2016flj,CMS:2016mah,ATLAS:2018xms}, is complicated due to effects like: sizeable regeneration or
recombination of quarkonium from the uncorrelated charm and anti-charm quark pairs in the QGP~\cite{Braun-Munzinger:2000csl,Thews:2000rj,Zhou:2014kka,He:2021zej}, cold nuclear matter effects~\cite{Vogt:2015uba} and feed-down contributions from the
decay of excited states. However, in the region of $p_\text{T}^{J/\psi}\gg m_{J/\psi}$ with $m_{J/\psi}$ the $J/\psi$ mass, the aforementioned dissociation and regeneration effects are negligible while jet quenching plays a dominant role for the $J/\psi$ suppression~\cite{Zhang:2022rby}. Recently, CMS Collaboration reported the measurement of the fragmentation of jets containing a prompt J$/\psi$ meson in PbPb and pp collisions at $\sqrt{s_\mathrm{NN}} =$ 5.02 TeV~\cite{CMS:2021puf}.
The measurement shows that the $J/\psi$ yield in jets is significantly suppressed in PbPb collisions comparing to that in pp, and the ratio of yield in PbPb to pp shows a rising trend with increasing $z_{J/\psi}$, different with the predicted pattern of longitudinal momentum fraction of $D^0$ in jets due to charm quark energy loss~\cite{Li:2022tcr}. This contrast questions the intuitive expectation that the nuclear modification for $J/\psi$ production is driven by its constituent charm quark. To the best of our knowledge, this phenomenon oberved by CMS collaboration has never been described and explained by any theoretical model calculations. 

The above observations motivate us to conduct a systematic study of the $J/\psi$ production in jet in both pp and AA collisions. Such a systematic study will not only be critical in our understanding of $J/\psi$ production mechanism, but also help to constrain the flavor dependent jet quenching mechanism~\cite{Zhang:2022rby}. In pp collisions, we utilize a theoretical framework analogous to the gluon fragmentation improved PYTHIA method~\cite{Bain:2017wvk,Bain:2016clc}, where we convolve hard QCD partonic cross sections with leading order NRQCD $J/\psi$ 
fragmentation functions. The parton propagation in QGP medium is simulated by the Linear Boltzmann Transport (LBT) model~\cite{ He:2015pra, Cao:2016gvr}
with bulk medium evolution provided by the CLVisc 3+1D hydrodynamic~\cite{Pang:2012he}.
The results are compared with available data in pp and PbPb collisions ~\cite{CMS:2021puf,LHCb:2017llq,Yang:2021smr}. In pp collisions, we demonstrate that the $z_{J/\psi}$ distribution is dominated by the gluon fragmentation~\cite{Zhang:2022rby}. In PbPb collisions, we find that the nuclear modification factor $R_{\text{AA}}(z_{J/\psi})$ is mainly driven by the gluon energy loss effect. Furthermore, we provide predictions for the jet transverse momentum and jet radius dependence of $z_{J/\psi}$ distribution and  nuclear modification factor $R_{\text{AA}}(z_{J/\psi})$, which can be tested in future experimental measurements.

The remainder of the paper is organized as follows. We first introduce the theoretical framework and present numerical results for $J/\psi$ production in jet in pp collisions in Sec.~\ref{sec:nume}. Based on the pp baseline, we present results for nuclear modification on $z_{J/\psi}$ distributions in PbPb collisions in Sec.~\ref{sec:nume}. Finally, A summary is given in Sec.\ref{summary}.

\section{$J/\psi$ production in jet in pp collisions}
\label{sec:Fra}
\subsection{Theoretical framework}
We start this section by introducing the theoretical framework suitable for describing high transverse momentum $J/\psi$ production in pp collisions. Within the framework of collinear factorization formalism \cite{Collins:1981uw}, the distribution of $J/\psi$ meson within a fully reconstructed jet in pp collisions can be written into the following schematic form \cite{Kaufmann:2015hma,Kang:2017yde} 
\begin{equation}
\label{pp-fac}
d\sigma[pp \to ({\rm jet} ~ J/\psi)+X]
=\sum_i d \hat \sigma_{pp \to ({\rm jet} ~ i)+X} \otimes D_{i\rightarrow J/\psi} ,
\end{equation}
where $\otimes$ denotes convolution product over the partonic momentum fraction, $\sum_i$ represents the sum over all relevant partonic channels. $d \hat \sigma_{pp \to ({\rm jet} ~ i)+X}$ is the differential cross section for inclusive parton $i$ distribution inside a jet in pp collisions, and $D_{i\rightarrow J/\psi}$ are the fragmentation functions (FFs) describing the transition of parton $i$ into the final observed $J/\psi$. In our calculation, we simulate the differential cross section $d \hat \sigma_{pp \to ({\rm jet} ~ i)+X}$ with parton $i$ generated by MadGraph \cite{Alwall:2014hca} and showered using PYTHIA \cite{Sjostrand:2014zea} down to $2m_\text{c}$ to obtain the transverse momentum $p_\text{T}^{\text{jet}}$ distribution for the measured jet. We use the parton distribution function (PDF) set ``CT14NLO" \cite{Dulat:2015mca} in LHAPDF \cite{Buckley:2014ana} to provide the information for partons that participate in the hard interaction in pp collisions. 
The $J/\psi$ FF can be further factorized at an initial scale $\mu_0 \sim m_{J/\psi}$ within the framework of NRQCD \cite{Ma:2013yla}
\begin{equation}
D_{i\rightarrow J/\psi}(z,\mu_0) = \sum_n \hat{d}_{i\to [c\bar c(n)]}(z,\mu_0) \langle \mathcal{O}_{[c\bar c(n)]}^{J/\psi}\rangle,
\end{equation}
where $\sum_n$ represents the sum over all possible intermediate nonrelativistic $c\bar c$ states, and the quantum number $n$ takes the standard notation as $n=~^{2S+1}L^{[i]}_J$ with the superscripts $i = 1$ and $8$ denote color-singlet and color-octet states, respectively. $\hat{d}_{i\to [c\bar c(n)]}$ are the short distance coefficients for the conversion of parton $i$ to intermediate nonrelativistic $c\bar c(n)$ quantum state, and have been derived, see, e.g., in Refs. \cite{Ma:2013yla,Bodwin:2003wh}. $z$ is the longitudinal momentum fraction of parton $i$ carried by the $c\bar c$ pair. 
$\langle \mathcal{O}_{[c\bar c(n)]}^{J/\psi}\rangle$ are the nonperturbative NRQCD long distance matrix elements (LDMEs), describing the transition of a $c\bar{c}$ pair in the quantum state $n$ into a final observed $J/\psi$. In our simulation, the NRQCD fragmentation functions are evaluated at leading order in QCD.

For $J/\psi$ production, we consider the most important four relevant states: $^3S^{[1]}_1$, $^1S^{[8]}_0$, $^3S^{[8]}_1$, $^3P^{[8]}_J$. The corresponding LDMEs have been extracted from experimental data by different groups, see, e.g., ~\cite{Bodwin:2014gia,Bodwin:2015iua,Gong:2012ug,Chao:2012iv,Ma:2010jj,Brambilla:2022ayc}. Specifically we take the results from fits to prompt $J/\psi$ production at large transverse momentum~\cite{Bodwin:2015iua}, which are more relevant to our case with high $p_\text{T}^{\rm jet}$ jet production. Notice that the fragmentation to $J/\psi$ from light quarks are suppressed by two orders of magnitude comparing to the others ~\cite{Baumgart:2014upa}. Therefore, only the contributions from charm and gluon fragmentation are considered in our paper. 

\subsection{Phenomenology at RHIC and the LHC energies}
In order to compare with the experimental data, we select $J/\psi$ and the associated jet according to
the kinematic cuts adopted by RHIC and the LHC measurements~\cite{CMS:2021puf,LHCb:2017llq,Yang:2021smr}, as detailed in Table.~\ref{table:cuts}. We implement the muon cuts by
assuming the $J/\psi$ are unpolarized and, therefore, decays
to dilepton isotropically in its rest frame.  The kinematic cut on  dilepton  clearly suppresses contributions from partons with low $z$ and, hence, reduce the contribution from soft
quark/gluon initiated $J/\psi$.

\begin{table}[!t]

\begin{center}
\caption{The kinematic cuts applied to $J/\psi$ in jet measurements at RHIC and the LHC.  }
  \begin{tabular}{p{1.0cm}<{\centering}|p{2.4cm}<{\centering}|p{2.4cm}<{\centering}|p{2.4cm}<{\centering}}
   \hline
               &   LHCb~\cite{LHCb:2017llq}       &     CMS~\cite{CMS:2021puf}        &  STAR~\cite{Yang:2021smr}    \\ \hline
    $\sqrt{s}$ &    13 TeV    &     5.02 TeV   &  500 GeV   \\ \hline
      Jet      &  $R=0.5$, \ \ \ \   $p_\text{T}^{\text{jet}}>20$ GeV/c, $2.5<\eta^{\text{jet}}<4.0$  &  $R=0.3$, $30<p_\text{T}^{\text{jet}}<40$ GeV/c, $|\eta^{\text{jet}}|<2.0$ & $R=0.4$,\ \ \  $p_\text{T}^{\text{jet}}>10$ GeV/c, $|\eta^{\text{jet}}|<0.6$  \\  \hline
     $J/\psi$      & $p(\mu)>5$ GeV/c, $2.5<\eta^{\mu}<4.0$    & $p_\text{T}^{J/\psi}>6.5$ GeV/c & $p_\text{T}^{J/\psi}>5$ GeV/c  \\  \hline
  \end{tabular}

  \label{table:cuts}
\end{center}
\vspace{-10pt}
\end{table}

\begin{figure}
  \centering
  \includegraphics[width=0.45\textwidth]{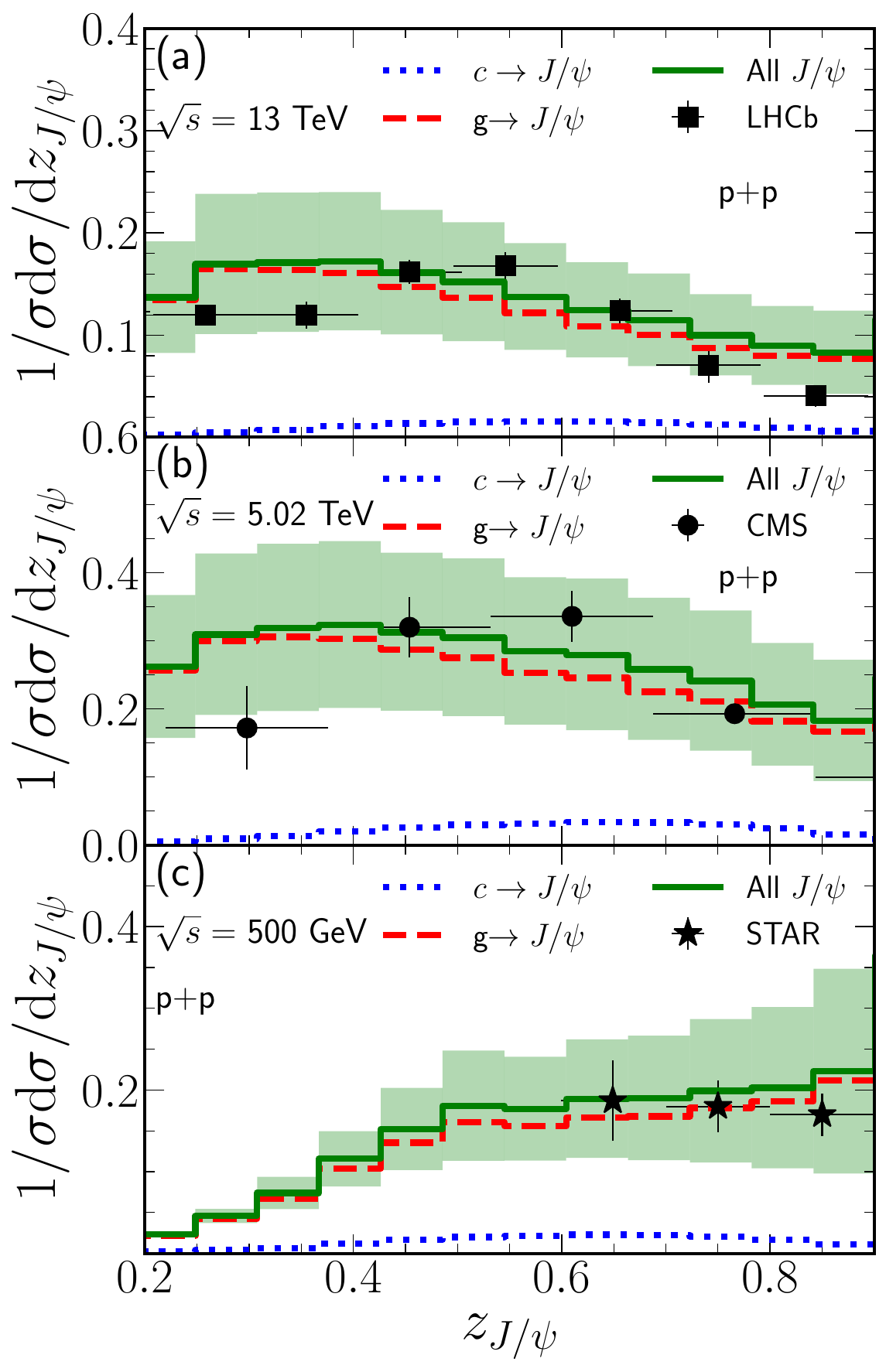}
  \caption{(Color online) Distributions of $J/\psi$  in jet (green line)  as well as from  gluon (red dashed line) and charm quark (solid blue line) fragmentation calculated as function of $z_{J/\psi}= p_\text{T}^{J/\psi}/p_\text{T}^{\text{jet}}$ in pp collision at $\sqrt{s}$:  (a) 13 TeV, (b) 5.02 TeV, and (c) 500 GeV.  The experimental data are taken from Refs. \cite{CMS:2021puf,LHCb:2017llq,Yang:2021smr}.
    }\label{frag_pp_gc}
\end{figure}

In Fig.~\ref{frag_pp_gc}, we show the results for jet fragmentation function for $J/\psi$ production in jet as a function of $z_{J/\psi}$ in pp collision at three different collision energies: (a) 13 TeV, (b) 5.02 TeV, and (c) 500 GeV, as well as the comparison with experimental data from LHCb~\cite{LHCb:2017llq}, CMS~\cite{CMS:2021puf}, and STAR~\cite{Yang:2021smr}, respectively. In order to satisfy the kinematic cuts and ensure the stability of the evaluated cross section at $z_{J/\psi}\rightarrow 1$, we limit our comparison with data in the range $0.2 < z_{J/\psi} < 0.9$ and normalize the $z_{J/\psi}$ distributions to the sum of the data in these bins as in Ref.~\cite{Bain:2017wvk}.
Our numerical result, as shown by the green band in Fig.~\ref{frag_pp_gc}, can quantitatively describe the experimental data within uncertainties from LDMEs (mainly comes from the counterbalance of  $c\bar{c}(^3S^{[8]}_1)$ and $c\bar{c}(^3P^{[8]}_J)$ channel).
The contributions from gluon (red dashed lines) and charm (blue dotted lines) fragmentation are also shown to demonstrate the $J/\psi$ production mechanism. One can see clearly that $J/\psi$ in jet is driven by gluon fragmentation ($>85\%$) in a wide range of $z_{J/\psi}$, while charm fragmentation is rather limited, in agreements with the statements in Refs~.\cite{Bain:2017wvk,CMS:2019ebt,Zhang:2022rby} for inclusive $J/\psi$ production. It is worth to point out that the experimental errors are generally smaller than the theoretical uncertainties, such a fact indicates the power of using jet fragmentation function to further constrain the LDMEs, similar to that for light hadron \cite{Gao:2024nkz} and D-meson FFs \cite{Anderle:2017cgl}.   

It is interesting to note that the $z_{J/\psi}$ distributions are different in three collision energies, with the peak of $z_{J/\psi}$ shifts to smaller values from RHIC to LHC energies. This phenomenon can be understood by the fact that the parton shower at RHIC energy is weaker than that at LHC energies, therefore less partons are produced within jet, which eventually lead to harder momentum distributions of charm and gluon in jet at RHIC energy. Combined with the fact that the gluon fragmentation function peaks at $z \sim 1$, we expect harder $z_{J/\psi}$ distribution at lower collision energy, which explains the energy dependence of peak position for $z_{J/\psi}$ distribution shown in Fig.~\ref{frag_pp_gc}.

\begin{figure}
  \centering
  \includegraphics[width=0.4\textwidth]{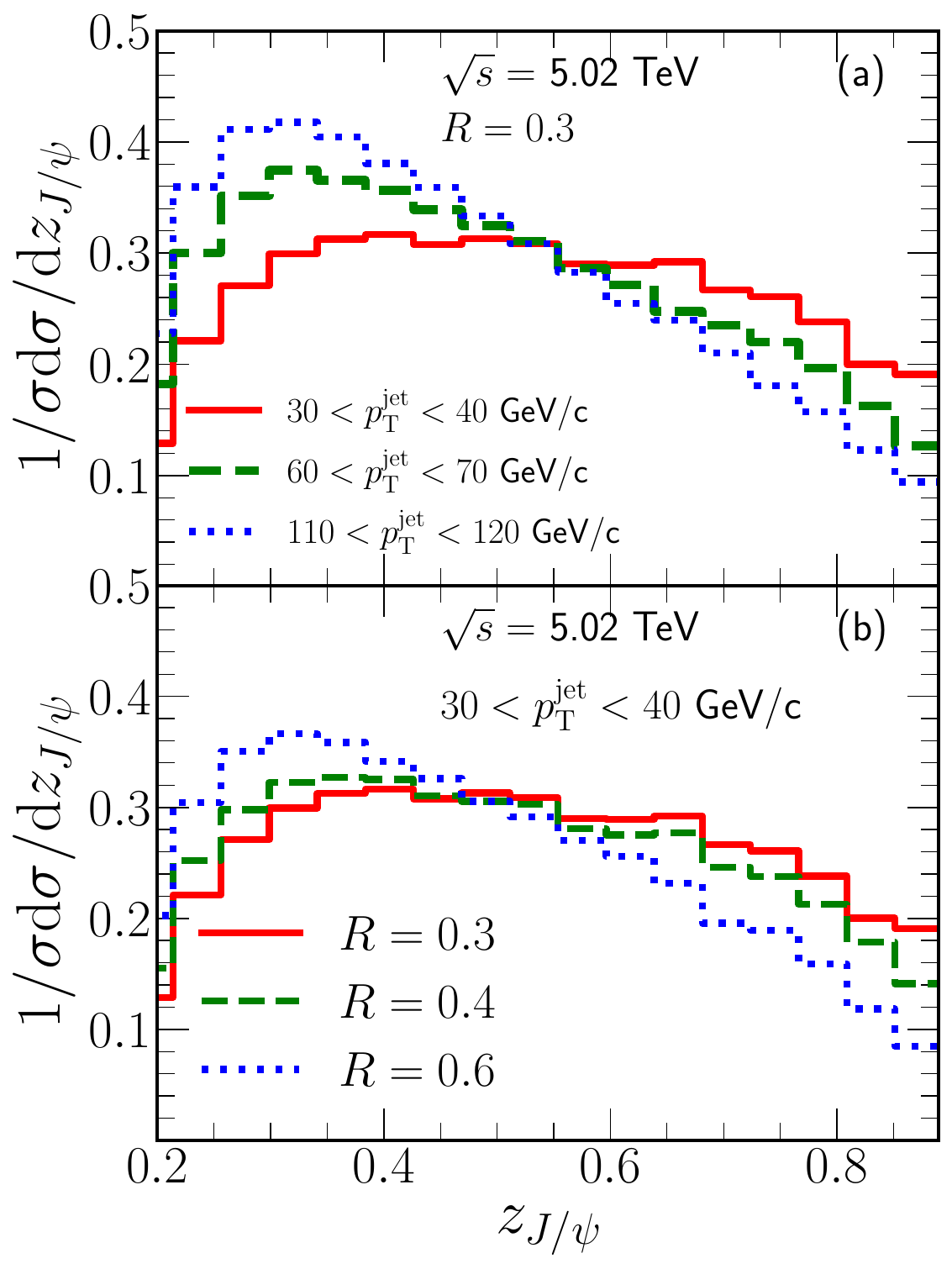}
  \caption{(Color online) The normalized distributions of $J/\psi$ in jet as a function of $z_{J/\psi}$ in: (a) three $p_\text{T}^{\text{jet}}$ intervals with jet cone $R=0.3$,  (b) $30<p_\text{T}^{\text{jet}}<40$ GeV/c  interval with three jet cone R =0.2, 0.4,  and 0.6, in p+p collisions  at $\sqrt{s}= 5.02 $ TeV.  }\label{frag_R1}
\end{figure}
The mechanism mentioned above can be tested by a more differential study in terms of jet transverse momentum $p_\text{T}^{\text{jet}}$ and jet radius $R$ dependencies of $z_{J/\psi}$ distribution. In Fig.~\ref{frag_R1}(a), we present predictions for $J/\psi$ production within jet in pp collisions as a function of $z_{J/\psi}$ at $\sqrt{s}= 5.02 $ TeV in three different $p_\text{T}^{\text{jet}}$ and $R$ intervals. The distributions are normalized to unity in $0.3<z_{J/\psi}<0.8$. As can be seen from the top panel that the $z_{J/\psi}$ is shifted to softer regions with the increase of $p_\text{T}^{\text{jet}}$, which is qualitatively consistent with the CMS measurements~\cite{CMS:2019ebt}.
This can be understood in a similar fashion as in Fig.~\ref{frag_pp_gc}, i.e., more gluons is generated in larger $p_\text{T}^{\text{jet}}$ region, therefore softer momentum distributions with the increase of $p_\text{T}^{\text{jet}}$.
We show in Fig. \ref{frag_R1}(b) the $R$ dependence in $30<p_\text{T}^{\text{jet}}<40$ GeV/c with three jet radius $R=0.3, 0.4, 0.6$. As we expect that larger size of jet contains more particles because of more radiated gluons are restored, the peak value of $z_{J/\psi}$ distribution shifts to smaller value with larger jet cone $R$. Such predicted $p_\text{T}^{\text{jet}}$ and $R$ dependent distributions can be tested in future measurements to confirm the production mechanism of $J/\psi$ in jet.

\section{$J/\psi$ production in jet in AA collisions}
\label{sec:nume}

\subsection{Framework for jet propagation in medium}
In relativistic heavy ion collisions, the energetic partons from hard interactions encounter multiple scatterings inside the QGP medium before they fragment into high-$p_\text{T}$ heavy quarkonium. We use the well developed Linear Boltzmann Transport (LBT) model to incorporate both elastic and inelastic processes for the charm quarks and gluons scattering with medium constituents~\cite{ He:2015pra, Cao:2016gvr}. For elastic scatterings, the evolution of hard partons are simulated by the following linear Boltzmann transport equation,
\begin{eqnarray}
&p_1\cdot\partial f_a(p_1)=-\int\frac{d^3p_2}{(2\pi)^32E_2}\int\frac{d^3p_3}{(2\pi)^32E_3}\int\frac{d^3p_4}{(2\pi)^32E_4} \nonumber \\
&\frac{1}{2}\sum _{b(c,d)}[f_a(p_1)f_b(p_2)-f_c(p_3)f_d(p_4)]|M_{ab\rightarrow cd}|^2 \nonumber \\
&\times S_2(s,t,u)(2\pi)^4\delta^4(p_1+p_2-p_3-p_4) + {\rm inel.}
\end{eqnarray}
where $f_{i=a,b,c,d}$ are the phase-space distributions of jet shower partons ($a, c$) and medium partons ($b, d$), $|M_{ab\rightarrow cd}|$ are the corresponding elastic scattering matrix elements which are regulated by a Lorentz-invariant regulation condition $S_2(s,t,u)=\theta(s>2\mu^{2}_{D})\theta(-s+\mu^{2}_{D}\leq t \leq -\mu^{2}_{D})$, with $\mu_{D}^{2}=g^{2}T^{2}(N_{c}+N_{f}/2)/3$ the Debye screening mass. The effect of  inelastic scatterings is described by the higher-twist formalism for induced gluon radiation as follows~\cite{Guo:2000nz,Zhang:2003wk},
\begin{eqnarray}
\frac{dN_g}{dxdk_\perp^2 dt}=\frac{6\alpha_s P(x ) \hat{q} k_\perp^4}{\pi (k_\perp^2+x^2 M^2)^4} \sin^2\left(\frac{t-t_i}{2\tau_f}\right),
\end{eqnarray}
Here $x$ denotes the energy fraction of the radiated gluon relative to a parent parton with mass $M$, $k_\perp$ is the transverse momentum. A lower energy cut-off $x_{min}=\mu_{D}/E$ is applied for the emitted gluon in the calculation. $P(x)$ is the splitting function in vacuum, and $\tau_f=2Ex(1-x)/(k^2_\perp+x^2M^2)$ is the formation time of the radiated gluons in QGP. The dynamic evolution of bulk medium is given by 3+1D CLVisc hydrodynamical model~\cite{Pang:2012he} with parameters fixed by reproducing hadron spectra from experimental measurements.
The details of LBT can be found in Ref.~\cite{Luo:2023nsi}, which has been implemented to successfully describe the suppression of experimental data on hadron, inclusive jets, $\gamma +$hadron/jets correlations and Z+jet production, as well as the prompt $J/\psi$ production \cite{Zhang:2022rby}. In our calculation, we follow previous study~\cite{Zhang:2018urd} to fix the free parameter $\alpha_s=0.18$ to control the strength of parton-medium interactions. 

\subsection{Numerical results}
\begin{figure}
  \centering
  \includegraphics[width=0.45\textwidth]{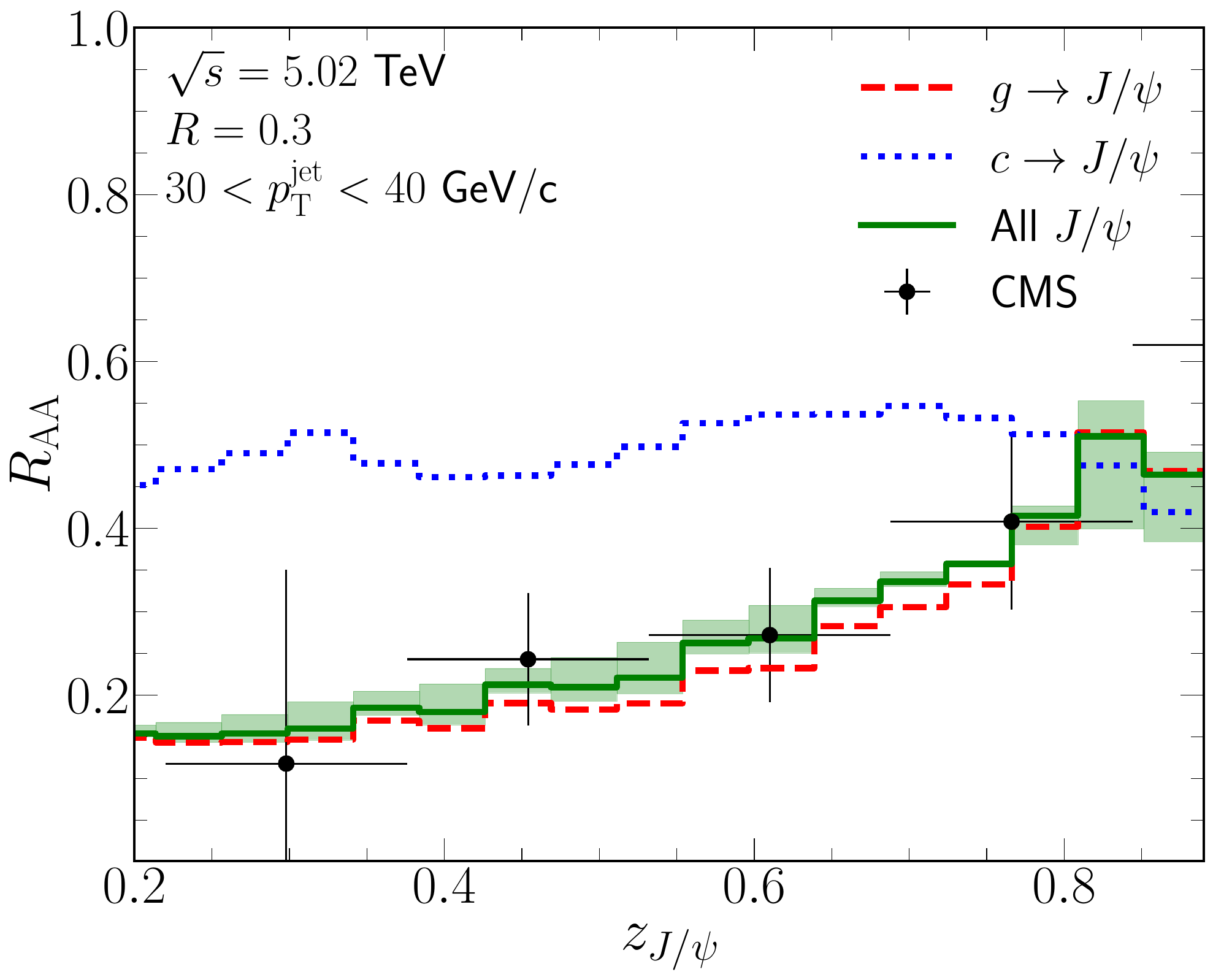}

  \caption{(Color online) The nuclear modification factor $R_{\text{AA}}$, as a function of $z_{J/\psi}$ at $\sqrt{s}= 5.02 $ TeV and the comparison with CMS measurement~\cite{CMS:2021puf}. The medium modifications on the $J/\psi$ cross section from gluon (red) and charm (blue) fragmentation are also shown for illustration.}\label{frag_raa}
\end{figure}

In Fig.~\ref{frag_raa}, we present the nuclear modification factor $R_{\text{AA}}$, defined as the ratio of $z_{J/\psi}$ distribution between  PbPb and pp collisions at $\sqrt{s}= 5.02 $ TeV, and compare with the CMS measurement \cite{CMS:2021puf}. The $z_{J/\psi}$ distribution is overall suppressed in PbPb collisions relative to that in pp collisions, and is quantitatively consistent with the experimental measurements within the uncertainties. The overall suppression has a natural interpretation in terms of the jet quenching phenomenon. The yields of jet and $J/\psi$ that pass all the selection cuts in PbPb collisions are reduced due to parton-medium interactions.  
The $z_{J/\psi}$ dependent $R_{\text{AA}}$ show a slight rising trend with increasing $z_{J/\psi}$, this is because of the reason that 
lower values of $z_{J/\psi}$ should be populated with jets with a $J/\psi$ produced late in the parton shower. Such a parton cascade is expected to have a large degree of interaction with the QGP in the form of subsequent medium-induced emissions~\cite{Caucal:2018dla},
as compared to a jet with a small partonic multiplicity. 

To illustrate the underlying mechanism for the suppression of $z_{J/\psi}$, we plot in Fig.~\ref{frag_raa} the medium modification on the $z_{J/\psi}$ from gluon and charm fragmentation as well. We see that $J/\psi$ from gluon fragmentation (red line) is much more suppressed compared to charm quark fragmentation (blue line) due to the parton mass and color-charge effect, where gluon lose much larger fraction of its energy in the medium compared to light-quark and $c$-quark jet as analyzed in \cite{Zhang:2022rby,Xing:2023ciw}. Considering the fact that gluon fragmentation dominates the $z_{J/\psi}$ distribution in pp collisions, an interesting observation is that the strong suppression of high-$p_\text{T}$ $J/\psi$ production in AA collisions is mainly driven by the gluon jet quenching.
Besides, as shown by the red line in Fig.~\ref{frag_raa}, the nuclear modification for gluon fragmentation also show a rising trend in $z_{J/\psi}$, while the one for charm quark represented by blue line show much weaker dependence on $z_{J/\psi}$. Those modification patterns 
will be explained in detail in the following.

To further clarify the medium modification of $z_{J/\psi}$ distribution, the relationship  between the fraction
variable $z$ in AA and pp collisions can be expressed as
\begin{equation}
  z^\text{AA}=\frac{p_\text{T,AA}^{J/\psi}}{p_\text{T,AA}^{\text{jet}}}= \frac{(1-f^{J/\psi})p_\text{T,pp}^{J/\psi}}{(1-f^{\text{jet}})p_\text{T,pp}^{\text{jet}}}=\frac{(1-f^{J/\psi})}{(1-f^{\text{jet}})}z^\text{pp},
\end{equation}
where $z^\text{AA}$ and $z^\text{pp}$ are the fraction variables after and before jet quenching, respectively.  $f^{J/\psi}$ and $f^{\text{jet}}$ are energy loss fractions of $J/\psi$ and jet when propagating through the medium. In the large $z$ region, where the $c$ quark is produced with fewer associated partons, the radiated gluon from charm quark may still fall in the jet cone and lead to smaller energy loss comparing to charm quark itself. As a consequence, we will have $f^{J/\psi}\geq f^{\text{jet}}$, thus find $z^\text{AA}<z^\text{pp}$ for $c$ quark fragmentation, similar with the modification pattern of $D^0$ in jet~\cite{Li:2022tcr}. Whereas for gluon fragmentation, jet has a larger number of sources of energy loss and tend to lose more energy~\cite{Pablos:2019ngg}, based on which,  $f^{J/\psi}$ will be smaller than $f^{\text{jet}}$, then gives $z^\text{AA}>z^\text{pp}$. This analysis is qualitatively in consistent with our numerical results, which 
demonstrate the role of gluon jet quenching in driving the nuclear suppression of $J/\psi$ in jet. 

\begin{figure}
  \centering
  \includegraphics[width=0.45\textwidth]{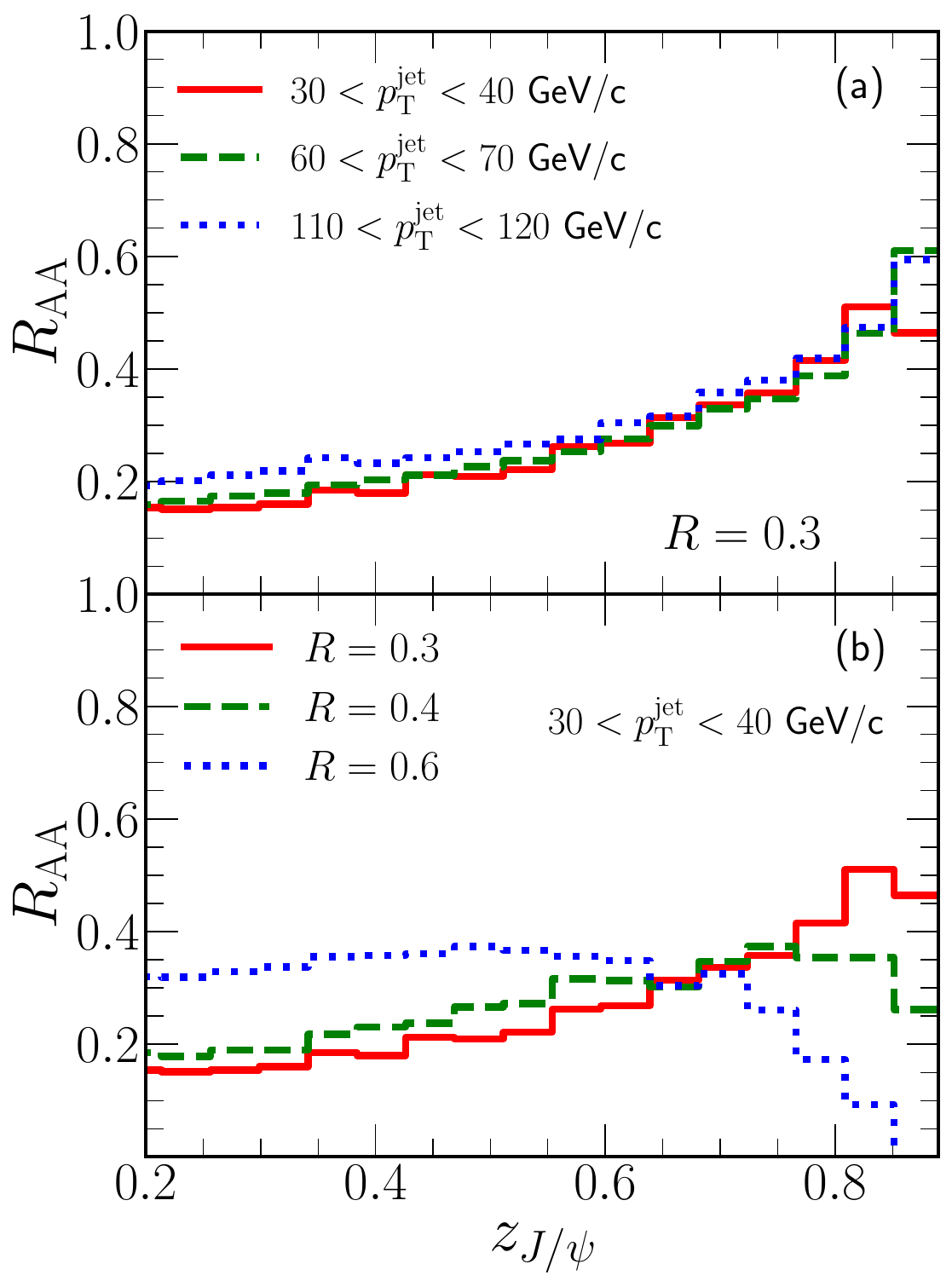}
  \caption{(Color online)  The nuclear modification factor $R_{\text{AA}}$  evaluated as a function of $z_{J/\psi}$  (a) in three $p_\text{T}^{\text{jet}}$ intervals with jet cone $R=0.3$,  (b) in $30<p_\text{T}^{\text{jet}}<40$ GeV/c  interval with three jet cone $R = $ 0.2, 0.4,  and 0.6 at $\sqrt{s}= 5.02 $ TeV. 
  }\label{frag_raa_R}
\end{figure}

In order to further test the mechanisms mentioned above. We present predictions in Fig. \ref{frag_raa_R} for the dependence of nuclear modification factor $R_{\text{AA}}$ on $p_\text{T}^{\text{jet}}$ and $R$ .
As shown in the top panel, we predict mild dependence of $R_{\text{AA}}$ on $p_\text{T}^{\text{jet}}$.
This can be explained by the fact that $f^\text{jet}$ and $f^{J/\psi}$ have similar downward tendency as shown in Refs.~\cite{Zhang:2023oid,Zhang:2022rby}. Therefore, the ratio between $f^{J/\psi}$ and $f^{\text{jet}}$ is  only moderately modified, leading to the predicted mild dependence on $p_\text{T}^{\text{jet}}$. Shown in the bottom panel is the prediction for nuclear modification factor $R_{\text{AA}}$ with three jet radius over the $p_\text{T}$ interval of [30, 40] GeV/c.
It turns out that the modification pattern of $R_{\text{AA}}(z_{J/\psi})$ is  significantly modified with increasing jet radius. For example, the $z_{J/\psi}$ dependent nuclear modification factor shows a downward (rising) trend at large $z_{J/\psi}$ with $R=0.6~(0.3)$.
This is driven by fact that larger fraction of radiations induced by the medium will be restored with larger jet radius $R$, thus leads to smaller energy loss fraction, i.e. $f^{\text{jet}}$ decrease with increasing jet radius $R$. Therefore, one would expect smaller $z^\text{AA}$ with $R=0.6$ comparing to $R=0.3$, leading to less suppression in small $z_{J/\psi}$ and more suppression in large $z_{J/\psi}$. 
However, we realize that the shape of $R_{\text{AA}}(z_{J/\psi})$ from charm fragmentation is hardly changed, simply because that 
charm jet lose much less fraction of its energy, leading to weaker jet radius dependence, while that from gluon fragmentation is  changed more significant. As the $J/\psi$ production is mainly driven by gluon fragmentation, we conclude that $R_{\text{AA}}(z_{J/\psi})$ should be dominated by the medium modification on gluon fragmentation processes. The upcoming high precision measurements of jet radius dependent $R_{\text{AA}}(z_{J/\psi})$  at LHC will put new insight into our understanding of $J/\psi$ production mechanism, and the charm and gluon jet quenching mechanism.

\section{Summary}
\label{summary}
In the paper, we have studied $J/\psi$ production within a jet in pp and their modification in heavy-ion collisions.
Within the framework of leading power NRQCD factorization formalism, we convolve the differential cross section of charm quarks and gluons in jet with leading order NRQCD $J/\psi$ fragmentation functions. Then a linear Boltzmann transport model combined with hydrodynamics are implemented for the simulation of jet-medium interaction in
nucleus-nucleus collisions.
We found that $J/\psi$ production in jet is mainly dominated by gluon dynamics, similar to that for inclusive $J/\psi$ suppression in heavy-ion collisions.
Furthermore, we present predictions for the dependence of $J/\psi$ production in jet on jet transverse momentum $p_\text{T}^{\text{jet}}$ and jet radius $R$ in both pp and AA collisions. We expect mild $p_\text{T}^{\text{jet}}$ dependence and strong jet radius $R$ dependence of the nuclear modification $R_{\text{AA}}$. 
These predictions can be tested in future high precision measurements on $J/\psi$ in jet at the LHC, which can help us to further pin down the $J/\psi$ production mechanism, as well as the flavor dependent jet quenching mechanism, ultimately, the properties of the QGP itself.


{\bf Acknowledgments:} We thank Yiannis Makris, Xiaozhi Bai and Qian Yang for helpful discussions. This research is supported by the National Natural Science Foundation of China with Project No. 12147131, 12035007, 12022512, and Guangdong Major Project of Basic and Applied Basic Research  with No. 22020B0301030008, 2022A1515010683.

\vspace*{-.6cm}

\bibliography{main.bbl}

\begin{thebibliography}{60}%
\makeatletter
\providecommand \@ifxundefined [1]{%
 \@ifx{#1\undefined}
}%
\providecommand \@ifnum [1]{%
 \ifnum #1\expandafter \@firstoftwo
 \else \expandafter \@secondoftwo
 \fi
}%
\providecommand \@ifx [1]{%
 \ifx #1\expandafter \@firstoftwo
 \else \expandafter \@secondoftwo
 \fi
}%
\providecommand \natexlab [1]{#1}%
\providecommand \enquote  [1]{``#1''}%
\providecommand \bibnamefont  [1]{#1}%
\providecommand \bibfnamefont [1]{#1}%
\providecommand \citenamefont [1]{#1}%
\providecommand \href@noop [0]{\@secondoftwo}%
\providecommand \href [0]{\begingroup \@sanitize@url \@href}%
\providecommand \@href[1]{\@@startlink{#1}\@@href}%
\providecommand \@@href[1]{\endgroup#1\@@endlink}%
\providecommand \@sanitize@url [0]{\catcode `\\12\catcode `\$12\catcode
  `\&12\catcode `\#12\catcode `\^12\catcode `\_12\catcode `\%12\relax}%
\providecommand \@@startlink[1]{}%
\providecommand \@@endlink[0]{}%
\providecommand \url  [0]{\begingroup\@sanitize@url \@url }%
\providecommand \@url [1]{\endgroup\@href {#1}{\urlprefix }}%
\providecommand \urlprefix  [0]{URL }%
\providecommand \Eprint [0]{\href }%
\providecommand \doibase [0]{http://dx.doi.org/}%
\providecommand \selectlanguage [0]{\@gobble}%
\providecommand \bibinfo  [0]{\@secondoftwo}%
\providecommand \bibfield  [0]{\@secondoftwo}%
\providecommand \translation [1]{[#1]}%
\providecommand \BibitemOpen [0]{}%
\providecommand \bibitemStop [0]{}%
\providecommand \bibitemNoStop [0]{.\EOS\space}%
\providecommand \EOS [0]{\spacefactor3000\relax}%
\providecommand \BibitemShut  [1]{\csname bibitem#1\endcsname}%
\let\auto@bib@innerbib\@empty
\bibitem [{\citenamefont {Andronic}\ \emph {et~al.}(2016)\citenamefont
  {Andronic} \emph {et~al.}}]{Andronic:2015wma}%
  \BibitemOpen
  \bibfield  {author} {\bibinfo {author} {\bibfnamefont {A.}~\bibnamefont
  {Andronic}} \emph {et~al.},\ }\href {\doibase 10.1140/epjc/s10052-015-3819-5}
  {\bibfield  {journal} {\bibinfo  {journal} {Eur. Phys. J. C}\ }\textbf
  {\bibinfo {volume} {76}},\ \bibinfo {pages} {107} (\bibinfo {year} {2016})},\
  \Eprint {http://arxiv.org/abs/1506.03981} {arXiv:1506.03981 [nucl-ex]}
  \BibitemShut {NoStop}%
\bibitem [{\citenamefont {Chapon}\ \emph {et~al.}(2022)\citenamefont {Chapon}
  \emph {et~al.}}]{Chapon:2020heu}%
  \BibitemOpen
  \bibfield  {author} {\bibinfo {author} {\bibfnamefont {E.}~\bibnamefont
  {Chapon}} \emph {et~al.},\ }\href {\doibase 10.1016/j.ppnp.2021.103906}
  {\bibfield  {journal} {\bibinfo  {journal} {Prog. Part. Nucl. Phys.}\
  }\textbf {\bibinfo {volume} {122}},\ \bibinfo {pages} {103906} (\bibinfo
  {year} {2022})},\ \Eprint {http://arxiv.org/abs/2012.14161} {arXiv:2012.14161
  [hep-ph]} \BibitemShut {NoStop}%
\bibitem [{\citenamefont {Lansberg}(2020)}]{Lansberg:2019adr}%
  \BibitemOpen
  \bibfield  {author} {\bibinfo {author} {\bibfnamefont {J.-P.}\ \bibnamefont
  {Lansberg}},\ }\href {\doibase 10.1016/j.physrep.2020.08.007} {\bibfield
  {journal} {\bibinfo  {journal} {Phys. Rept.}\ }\textbf {\bibinfo {volume}
  {889}},\ \bibinfo {pages} {1} (\bibinfo {year} {2020})},\ \Eprint
  {http://arxiv.org/abs/1903.09185} {arXiv:1903.09185 [hep-ph]} \BibitemShut
  {NoStop}%
\bibitem [{\citenamefont {Matsui}\ and\ \citenamefont
  {Satz}(1986)}]{Matsui:1986dk}%
  \BibitemOpen
  \bibfield  {author} {\bibinfo {author} {\bibfnamefont {T.}~\bibnamefont
  {Matsui}}\ and\ \bibinfo {author} {\bibfnamefont {H.}~\bibnamefont {Satz}},\
  }\href {\doibase 10.1016/0370-2693(86)91404-8} {\bibfield  {journal}
  {\bibinfo  {journal} {Phys. Lett. B}\ }\textbf {\bibinfo {volume} {178}},\
  \bibinfo {pages} {416} (\bibinfo {year} {1986})}\BibitemShut {NoStop}%
\bibitem [{\citenamefont {Flore}\ \emph {et~al.}(2020)\citenamefont {Flore},
  \citenamefont {Lansberg}, \citenamefont {Shao},\ and\ \citenamefont
  {Yedelkina}}]{Flore:2020jau}%
  \BibitemOpen
  \bibfield  {author} {\bibinfo {author} {\bibfnamefont {C.}~\bibnamefont
  {Flore}}, \bibinfo {author} {\bibfnamefont {J.-P.}\ \bibnamefont {Lansberg}},
  \bibinfo {author} {\bibfnamefont {H.-S.}\ \bibnamefont {Shao}}, \ and\
  \bibinfo {author} {\bibfnamefont {Y.}~\bibnamefont {Yedelkina}},\ }\href
  {\doibase 10.1016/j.physletb.2020.135926} {\bibfield  {journal} {\bibinfo
  {journal} {Phys. Lett. B}\ }\textbf {\bibinfo {volume} {811}},\ \bibinfo
  {pages} {135926} (\bibinfo {year} {2020})},\ \Eprint
  {http://arxiv.org/abs/2009.08264} {arXiv:2009.08264 [hep-ph]} \BibitemShut
  {NoStop}%
\bibitem [{\citenamefont {D'Alesio}\ \emph {et~al.}(2022)\citenamefont
  {D'Alesio}, \citenamefont {Maxia}, \citenamefont {Murgia}, \citenamefont
  {Pisano},\ and\ \citenamefont {Rajesh}}]{DAlesio:2021yws}%
  \BibitemOpen
  \bibfield  {author} {\bibinfo {author} {\bibfnamefont {U.}~\bibnamefont
  {D'Alesio}}, \bibinfo {author} {\bibfnamefont {L.}~\bibnamefont {Maxia}},
  \bibinfo {author} {\bibfnamefont {F.}~\bibnamefont {Murgia}}, \bibinfo
  {author} {\bibfnamefont {C.}~\bibnamefont {Pisano}}, \ and\ \bibinfo {author}
  {\bibfnamefont {S.}~\bibnamefont {Rajesh}},\ }\href {\doibase
  10.1007/JHEP03(2022)037} {\bibfield  {journal} {\bibinfo  {journal} {JHEP}\
  }\textbf {\bibinfo {volume} {03}},\ \bibinfo {pages} {037} (\bibinfo {year}
  {2022})},\ \Eprint {http://arxiv.org/abs/2110.07529} {arXiv:2110.07529
  [hep-ph]} \BibitemShut {NoStop}%
\bibitem [{\citenamefont {Boer}\ \emph {et~al.}(2021)\citenamefont {Boer},
  \citenamefont {Pisano},\ and\ \citenamefont {Taels}}]{Boer:2021ehu}%
  \BibitemOpen
  \bibfield  {author} {\bibinfo {author} {\bibfnamefont {D.}~\bibnamefont
  {Boer}}, \bibinfo {author} {\bibfnamefont {C.}~\bibnamefont {Pisano}}, \ and\
  \bibinfo {author} {\bibfnamefont {P.}~\bibnamefont {Taels}},\ }\href
  {\doibase 10.1103/PhysRevD.103.074012} {\bibfield  {journal} {\bibinfo
  {journal} {Phys. Rev. D}\ }\textbf {\bibinfo {volume} {103}},\ \bibinfo
  {pages} {074012} (\bibinfo {year} {2021})},\ \Eprint
  {http://arxiv.org/abs/2102.00003} {arXiv:2102.00003 [hep-ph]} \BibitemShut
  {NoStop}%
\bibitem [{\citenamefont {Chen}\ \emph {et~al.}(2023)\citenamefont {Chen},
  \citenamefont {Xing},\ and\ \citenamefont {Yoshida}}]{Chen:2023hvu}%
  \BibitemOpen
  \bibfield  {author} {\bibinfo {author} {\bibfnamefont {L.}~\bibnamefont
  {Chen}}, \bibinfo {author} {\bibfnamefont {H.}~\bibnamefont {Xing}}, \ and\
  \bibinfo {author} {\bibfnamefont {S.}~\bibnamefont {Yoshida}},\ }\href
  {\doibase 10.1103/PhysRevD.108.054021} {\bibfield  {journal} {\bibinfo
  {journal} {Phys. Rev. D}\ }\textbf {\bibinfo {volume} {108}},\ \bibinfo
  {pages} {054021} (\bibinfo {year} {2023})},\ \Eprint
  {http://arxiv.org/abs/2306.12647} {arXiv:2306.12647 [hep-ph]} \BibitemShut
  {NoStop}%
\bibitem [{\citenamefont {Qiu}\ \emph {et~al.}(2021)\citenamefont {Qiu},
  \citenamefont {Wang},\ and\ \citenamefont {Xing}}]{Qiu:2020xum}%
  \BibitemOpen
  \bibfield  {author} {\bibinfo {author} {\bibfnamefont {J.-W.}\ \bibnamefont
  {Qiu}}, \bibinfo {author} {\bibfnamefont {X.-P.}\ \bibnamefont {Wang}}, \
  and\ \bibinfo {author} {\bibfnamefont {H.}~\bibnamefont {Xing}},\ }\href
  {\doibase 10.1088/0256-307X/38/4/041201} {\bibfield  {journal} {\bibinfo
  {journal} {Chin. Phys. Lett.}\ }\textbf {\bibinfo {volume} {38}},\ \bibinfo
  {pages} {041201} (\bibinfo {year} {2021})},\ \Eprint
  {http://arxiv.org/abs/2005.10832} {arXiv:2005.10832 [hep-ph]} \BibitemShut
  {NoStop}%
\bibitem [{\citenamefont {Bodwin}\ \emph {et~al.}(1995)\citenamefont {Bodwin},
  \citenamefont {Braaten},\ and\ \citenamefont {Lepage}}]{Bodwin:1994jh}%
  \BibitemOpen
  \bibfield  {author} {\bibinfo {author} {\bibfnamefont {G.~T.}\ \bibnamefont
  {Bodwin}}, \bibinfo {author} {\bibfnamefont {E.}~\bibnamefont {Braaten}}, \
  and\ \bibinfo {author} {\bibfnamefont {G.~P.}\ \bibnamefont {Lepage}},\
  }\href {\doibase 10.1103/PhysRevD.55.5853} {\bibfield  {journal} {\bibinfo
  {journal} {Phys. Rev. D}\ }\textbf {\bibinfo {volume} {51}},\ \bibinfo
  {pages} {1125} (\bibinfo {year} {1995})},\ \bibinfo {note} {[Erratum:
  Phys.Rev.D 55, 5853 (1997)]},\ \Eprint {http://arxiv.org/abs/hep-ph/9407339}
  {arXiv:hep-ph/9407339} \BibitemShut {NoStop}%
\bibitem [{\citenamefont {Dai}\ and\ \citenamefont
  {Shrivastava}(2017)}]{Dai:2017cjq}%
  \BibitemOpen
  \bibfield  {author} {\bibinfo {author} {\bibfnamefont {L.}~\bibnamefont
  {Dai}}\ and\ \bibinfo {author} {\bibfnamefont {P.}~\bibnamefont
  {Shrivastava}},\ }\href {\doibase 10.1103/PhysRevD.96.036020} {\bibfield
  {journal} {\bibinfo  {journal} {Phys. Rev. D}\ }\textbf {\bibinfo {volume}
  {96}},\ \bibinfo {pages} {036020} (\bibinfo {year} {2017})},\ \Eprint
  {http://arxiv.org/abs/1707.08629} {arXiv:1707.08629 [hep-ph]} \BibitemShut
  {NoStop}%
\bibitem [{\citenamefont {Bain}\ \emph {et~al.}(2016)\citenamefont {Bain},
  \citenamefont {Dai}, \citenamefont {Hornig}, \citenamefont {Leibovich},
  \citenamefont {Makris},\ and\ \citenamefont {Mehen}}]{Bain:2016clc}%
  \BibitemOpen
  \bibfield  {author} {\bibinfo {author} {\bibfnamefont {R.}~\bibnamefont
  {Bain}}, \bibinfo {author} {\bibfnamefont {L.}~\bibnamefont {Dai}}, \bibinfo
  {author} {\bibfnamefont {A.}~\bibnamefont {Hornig}}, \bibinfo {author}
  {\bibfnamefont {A.~K.}\ \bibnamefont {Leibovich}}, \bibinfo {author}
  {\bibfnamefont {Y.}~\bibnamefont {Makris}}, \ and\ \bibinfo {author}
  {\bibfnamefont {T.}~\bibnamefont {Mehen}},\ }\href {\doibase
  10.1007/JHEP06(2016)121} {\bibfield  {journal} {\bibinfo  {journal} {JHEP}\
  }\textbf {\bibinfo {volume} {06}},\ \bibinfo {pages} {121} (\bibinfo {year}
  {2016})},\ \Eprint {http://arxiv.org/abs/1603.06981} {arXiv:1603.06981
  [hep-ph]} \BibitemShut {NoStop}%
\bibitem [{\citenamefont {Baumgart}\ \emph {et~al.}(2014)\citenamefont
  {Baumgart}, \citenamefont {Leibovich}, \citenamefont {Mehen},\ and\
  \citenamefont {Rothstein}}]{Baumgart:2014upa}%
  \BibitemOpen
  \bibfield  {author} {\bibinfo {author} {\bibfnamefont {M.}~\bibnamefont
  {Baumgart}}, \bibinfo {author} {\bibfnamefont {A.~K.}\ \bibnamefont
  {Leibovich}}, \bibinfo {author} {\bibfnamefont {T.}~\bibnamefont {Mehen}}, \
  and\ \bibinfo {author} {\bibfnamefont {I.~Z.}\ \bibnamefont {Rothstein}},\
  }\href {\doibase 10.1007/JHEP11(2014)003} {\bibfield  {journal} {\bibinfo
  {journal} {JHEP}\ }\textbf {\bibinfo {volume} {11}},\ \bibinfo {pages} {003}
  (\bibinfo {year} {2014})},\ \Eprint {http://arxiv.org/abs/1406.2295}
  {arXiv:1406.2295 [hep-ph]} \BibitemShut {NoStop}%
\bibitem [{\citenamefont {Kang}\ \emph {et~al.}(2017)\citenamefont {Kang},
  \citenamefont {Qiu}, \citenamefont {Ringer}, \citenamefont {Xing},\ and\
  \citenamefont {Zhang}}]{Kang:2017yde}%
  \BibitemOpen
  \bibfield  {author} {\bibinfo {author} {\bibfnamefont {Z.-B.}\ \bibnamefont
  {Kang}}, \bibinfo {author} {\bibfnamefont {J.-W.}\ \bibnamefont {Qiu}},
  \bibinfo {author} {\bibfnamefont {F.}~\bibnamefont {Ringer}}, \bibinfo
  {author} {\bibfnamefont {H.}~\bibnamefont {Xing}}, \ and\ \bibinfo {author}
  {\bibfnamefont {H.}~\bibnamefont {Zhang}},\ }\href {\doibase
  10.1103/PhysRevLett.119.032001} {\bibfield  {journal} {\bibinfo  {journal}
  {Phys. Rev. Lett.}\ }\textbf {\bibinfo {volume} {119}},\ \bibinfo {pages}
  {032001} (\bibinfo {year} {2017})},\ \Eprint
  {http://arxiv.org/abs/1702.03287} {arXiv:1702.03287 [hep-ph]} \BibitemShut
  {NoStop}%
\bibitem [{\citenamefont {Aaij}\ \emph {et~al.}(2017)\citenamefont {Aaij} \emph
  {et~al.}}]{LHCb:2017llq}%
  \BibitemOpen
  \bibfield  {author} {\bibinfo {author} {\bibfnamefont {R.}~\bibnamefont
  {Aaij}} \emph {et~al.} (\bibinfo {collaboration} {LHCb}),\ }\href {\doibase
  10.1103/PhysRevLett.118.192001} {\bibfield  {journal} {\bibinfo  {journal}
  {Phys. Rev. Lett.}\ }\textbf {\bibinfo {volume} {118}},\ \bibinfo {pages}
  {192001} (\bibinfo {year} {2017})},\ \Eprint
  {http://arxiv.org/abs/1701.05116} {arXiv:1701.05116 [hep-ex]} \BibitemShut
  {NoStop}%
\bibitem [{\citenamefont {Tumasyan}\ \emph {et~al.}(2022)\citenamefont
  {Tumasyan} \emph {et~al.}}]{CMS:2021puf}%
  \BibitemOpen
  \bibfield  {author} {\bibinfo {author} {\bibfnamefont {A.}~\bibnamefont
  {Tumasyan}} \emph {et~al.} (\bibinfo {collaboration} {CMS}),\ }\href
  {\doibase 10.1016/j.physletb.2021.136842} {\bibfield  {journal} {\bibinfo
  {journal} {Phys. Lett. B}\ }\textbf {\bibinfo {volume} {825}},\ \bibinfo
  {pages} {136842} (\bibinfo {year} {2022})},\ \Eprint
  {http://arxiv.org/abs/2106.13235} {arXiv:2106.13235 [hep-ex]} \BibitemShut
  {NoStop}%
\bibitem [{\citenamefont {Sirunyan}\ \emph {et~al.}(2020)\citenamefont
  {Sirunyan} \emph {et~al.}}]{CMS:2019ebt}%
  \BibitemOpen
  \bibfield  {author} {\bibinfo {author} {\bibfnamefont {A.~M.}\ \bibnamefont
  {Sirunyan}} \emph {et~al.} (\bibinfo {collaboration} {CMS}),\ }\href
  {\doibase 10.1016/j.physletb.2020.135409} {\bibfield  {journal} {\bibinfo
  {journal} {Phys. Lett. B}\ }\textbf {\bibinfo {volume} {804}},\ \bibinfo
  {pages} {135409} (\bibinfo {year} {2020})},\ \Eprint
  {http://arxiv.org/abs/1910.01686} {arXiv:1910.01686 [hep-ex]} \BibitemShut
  {NoStop}%
\bibitem [{\citenamefont {Yang}(2021)}]{Yang:2021smr}%
  \BibitemOpen
  \bibfield  {author} {\bibinfo {author} {\bibfnamefont {Q.}~\bibnamefont
  {Yang}} (\bibinfo {collaboration} {STAR}),\ }\href {\doibase
  10.22323/1.387.0072} {\bibfield  {journal} {\bibinfo  {journal} {PoS}\
  }\textbf {\bibinfo {volume} {HardProbes2020}},\ \bibinfo {pages} {072}
  (\bibinfo {year} {2021})}\BibitemShut {NoStop}%
\bibitem [{\citenamefont {Sj\"ostrand}\ \emph {et~al.}(2015)\citenamefont
  {Sj\"ostrand}, \citenamefont {Ask}, \citenamefont {Christiansen},
  \citenamefont {Corke}, \citenamefont {Desai}, \citenamefont {Ilten},
  \citenamefont {Mrenna}, \citenamefont {Prestel}, \citenamefont {Rasmussen},\
  and\ \citenamefont {Skands}}]{Sjostrand:2014zea}%
  \BibitemOpen
  \bibfield  {author} {\bibinfo {author} {\bibfnamefont {T.}~\bibnamefont
  {Sj\"ostrand}}, \bibinfo {author} {\bibfnamefont {S.}~\bibnamefont {Ask}},
  \bibinfo {author} {\bibfnamefont {J.~R.}\ \bibnamefont {Christiansen}},
  \bibinfo {author} {\bibfnamefont {R.}~\bibnamefont {Corke}}, \bibinfo
  {author} {\bibfnamefont {N.}~\bibnamefont {Desai}}, \bibinfo {author}
  {\bibfnamefont {P.}~\bibnamefont {Ilten}}, \bibinfo {author} {\bibfnamefont
  {S.}~\bibnamefont {Mrenna}}, \bibinfo {author} {\bibfnamefont
  {S.}~\bibnamefont {Prestel}}, \bibinfo {author} {\bibfnamefont {C.~O.}\
  \bibnamefont {Rasmussen}}, \ and\ \bibinfo {author} {\bibfnamefont {P.~Z.}\
  \bibnamefont {Skands}},\ }\href {\doibase 10.1016/j.cpc.2015.01.024}
  {\bibfield  {journal} {\bibinfo  {journal} {Comput. Phys. Commun.}\ }\textbf
  {\bibinfo {volume} {191}},\ \bibinfo {pages} {159} (\bibinfo {year}
  {2015})},\ \Eprint {http://arxiv.org/abs/1410.3012} {arXiv:1410.3012
  [hep-ph]} \BibitemShut {NoStop}%
\bibitem [{\citenamefont {Bain}\ \emph {et~al.}(2017)\citenamefont {Bain},
  \citenamefont {Dai}, \citenamefont {Leibovich}, \citenamefont {Makris},\ and\
  \citenamefont {Mehen}}]{Bain:2017wvk}%
  \BibitemOpen
  \bibfield  {author} {\bibinfo {author} {\bibfnamefont {R.}~\bibnamefont
  {Bain}}, \bibinfo {author} {\bibfnamefont {L.}~\bibnamefont {Dai}}, \bibinfo
  {author} {\bibfnamefont {A.}~\bibnamefont {Leibovich}}, \bibinfo {author}
  {\bibfnamefont {Y.}~\bibnamefont {Makris}}, \ and\ \bibinfo {author}
  {\bibfnamefont {T.}~\bibnamefont {Mehen}},\ }\href {\doibase
  10.1103/PhysRevLett.119.032002} {\bibfield  {journal} {\bibinfo  {journal}
  {Phys. Rev. Lett.}\ }\textbf {\bibinfo {volume} {119}},\ \bibinfo {pages}
  {032002} (\bibinfo {year} {2017})},\ \Eprint
  {http://arxiv.org/abs/1702.05525} {arXiv:1702.05525 [hep-ph]} \BibitemShut
  {NoStop}%
\bibitem [{\citenamefont {Adare}\ \emph {et~al.}(2007)\citenamefont {Adare}
  \emph {et~al.}}]{PHENIX:2006gsi}%
  \BibitemOpen
  \bibfield  {author} {\bibinfo {author} {\bibfnamefont {A.}~\bibnamefont
  {Adare}} \emph {et~al.} (\bibinfo {collaboration} {PHENIX}),\ }\href
  {\doibase 10.1103/PhysRevLett.98.232301} {\bibfield  {journal} {\bibinfo
  {journal} {Phys. Rev. Lett.}\ }\textbf {\bibinfo {volume} {98}},\ \bibinfo
  {pages} {232301} (\bibinfo {year} {2007})},\ \Eprint
  {http://arxiv.org/abs/nucl-ex/0611020} {arXiv:nucl-ex/0611020} \BibitemShut
  {NoStop}%
\bibitem [{\citenamefont {Adare}\ \emph {et~al.}(2011)\citenamefont {Adare}
  \emph {et~al.}}]{PHENIX:2011img}%
  \BibitemOpen
  \bibfield  {author} {\bibinfo {author} {\bibfnamefont {A.}~\bibnamefont
  {Adare}} \emph {et~al.} (\bibinfo {collaboration} {PHENIX}),\ }\href
  {\doibase 10.1103/PhysRevC.84.054912} {\bibfield  {journal} {\bibinfo
  {journal} {Phys. Rev. C}\ }\textbf {\bibinfo {volume} {84}},\ \bibinfo
  {pages} {054912} (\bibinfo {year} {2011})},\ \Eprint
  {http://arxiv.org/abs/1103.6269} {arXiv:1103.6269 [nucl-ex]} \BibitemShut
  {NoStop}%
\bibitem [{\citenamefont {Alessandro}\ \emph {et~al.}(2005)\citenamefont
  {Alessandro} \emph {et~al.}}]{NA50:2004sgj}%
  \BibitemOpen
  \bibfield  {author} {\bibinfo {author} {\bibfnamefont {B.}~\bibnamefont
  {Alessandro}} \emph {et~al.} (\bibinfo {collaboration} {NA50}),\ }\href
  {\doibase 10.1140/epjc/s2004-02107-9} {\bibfield  {journal} {\bibinfo
  {journal} {Eur. Phys. J. C}\ }\textbf {\bibinfo {volume} {39}},\ \bibinfo
  {pages} {335} (\bibinfo {year} {2005})},\ \Eprint
  {http://arxiv.org/abs/hep-ex/0412036} {arXiv:hep-ex/0412036} \BibitemShut
  {NoStop}%
\bibitem [{\citenamefont {Abelev}\ \emph {et~al.}(2012)\citenamefont {Abelev}
  \emph {et~al.}}]{ALICE:2012jsl}%
  \BibitemOpen
  \bibfield  {author} {\bibinfo {author} {\bibfnamefont {B.}~\bibnamefont
  {Abelev}} \emph {et~al.} (\bibinfo {collaboration} {ALICE}),\ }\href
  {\doibase 10.1103/PhysRevLett.109.072301} {\bibfield  {journal} {\bibinfo
  {journal} {Phys. Rev. Lett.}\ }\textbf {\bibinfo {volume} {109}},\ \bibinfo
  {pages} {072301} (\bibinfo {year} {2012})},\ \Eprint
  {http://arxiv.org/abs/1202.1383} {arXiv:1202.1383 [hep-ex]} \BibitemShut
  {NoStop}%
\bibitem [{\citenamefont {Adam}\ \emph {et~al.}(2017)\citenamefont {Adam} \emph
  {et~al.}}]{ALICE:2016flj}%
  \BibitemOpen
  \bibfield  {author} {\bibinfo {author} {\bibfnamefont {J.}~\bibnamefont
  {Adam}} \emph {et~al.} (\bibinfo {collaboration} {ALICE}),\ }\href {\doibase
  10.1016/j.physletb.2016.12.064} {\bibfield  {journal} {\bibinfo  {journal}
  {Phys. Lett. B}\ }\textbf {\bibinfo {volume} {766}},\ \bibinfo {pages} {212}
  (\bibinfo {year} {2017})},\ \Eprint {http://arxiv.org/abs/1606.08197}
  {arXiv:1606.08197 [nucl-ex]} \BibitemShut {NoStop}%
\bibitem [{\citenamefont {Khachatryan}\ \emph {et~al.}(2017)\citenamefont
  {Khachatryan} \emph {et~al.}}]{CMS:2016mah}%
  \BibitemOpen
  \bibfield  {author} {\bibinfo {author} {\bibfnamefont {V.}~\bibnamefont
  {Khachatryan}} \emph {et~al.} (\bibinfo {collaboration} {CMS}),\ }\href
  {\doibase 10.1140/epjc/s10052-017-4781-1} {\bibfield  {journal} {\bibinfo
  {journal} {Eur. Phys. J. C}\ }\textbf {\bibinfo {volume} {77}},\ \bibinfo
  {pages} {252} (\bibinfo {year} {2017})},\ \Eprint
  {http://arxiv.org/abs/1610.00613} {arXiv:1610.00613 [nucl-ex]} \BibitemShut
  {NoStop}%
\bibitem [{\citenamefont {Aaboud}\ \emph {et~al.}(2018)\citenamefont {Aaboud}
  \emph {et~al.}}]{ATLAS:2018xms}%
  \BibitemOpen
  \bibfield  {author} {\bibinfo {author} {\bibfnamefont {M.}~\bibnamefont
  {Aaboud}} \emph {et~al.} (\bibinfo {collaboration} {ATLAS}),\ }\href
  {\doibase 10.1140/epjc/s10052-018-6243-9} {\bibfield  {journal} {\bibinfo
  {journal} {Eur. Phys. J. C}\ }\textbf {\bibinfo {volume} {78}},\ \bibinfo
  {pages} {784} (\bibinfo {year} {2018})},\ \Eprint
  {http://arxiv.org/abs/1807.05198} {arXiv:1807.05198 [nucl-ex]} \BibitemShut
  {NoStop}%
\bibitem [{\citenamefont {Braun-Munzinger}\ and\ \citenamefont
  {Stachel}(2000)}]{Braun-Munzinger:2000csl}%
  \BibitemOpen
  \bibfield  {author} {\bibinfo {author} {\bibfnamefont {P.}~\bibnamefont
  {Braun-Munzinger}}\ and\ \bibinfo {author} {\bibfnamefont {J.}~\bibnamefont
  {Stachel}},\ }\href {\doibase 10.1016/S0370-2693(00)00991-6} {\bibfield
  {journal} {\bibinfo  {journal} {Phys. Lett. B}\ }\textbf {\bibinfo {volume}
  {490}},\ \bibinfo {pages} {196} (\bibinfo {year} {2000})},\ \Eprint
  {http://arxiv.org/abs/nucl-th/0007059} {arXiv:nucl-th/0007059} \BibitemShut
  {NoStop}%
\bibitem [{\citenamefont {Thews}\ \emph {et~al.}(2001)\citenamefont {Thews},
  \citenamefont {Schroedter},\ and\ \citenamefont {Rafelski}}]{Thews:2000rj}%
  \BibitemOpen
  \bibfield  {author} {\bibinfo {author} {\bibfnamefont {R.~L.}\ \bibnamefont
  {Thews}}, \bibinfo {author} {\bibfnamefont {M.}~\bibnamefont {Schroedter}}, \
  and\ \bibinfo {author} {\bibfnamefont {J.}~\bibnamefont {Rafelski}},\ }\href
  {\doibase 10.1103/PhysRevC.63.054905} {\bibfield  {journal} {\bibinfo
  {journal} {Phys. Rev. C}\ }\textbf {\bibinfo {volume} {63}},\ \bibinfo
  {pages} {054905} (\bibinfo {year} {2001})},\ \Eprint
  {http://arxiv.org/abs/hep-ph/0007323} {arXiv:hep-ph/0007323} \BibitemShut
  {NoStop}%
\bibitem [{\citenamefont {Zhou}\ \emph {et~al.}(2014)\citenamefont {Zhou},
  \citenamefont {Xu}, \citenamefont {Xu},\ and\ \citenamefont
  {Zhuang}}]{Zhou:2014kka}%
  \BibitemOpen
  \bibfield  {author} {\bibinfo {author} {\bibfnamefont {K.}~\bibnamefont
  {Zhou}}, \bibinfo {author} {\bibfnamefont {N.}~\bibnamefont {Xu}}, \bibinfo
  {author} {\bibfnamefont {Z.}~\bibnamefont {Xu}}, \ and\ \bibinfo {author}
  {\bibfnamefont {P.}~\bibnamefont {Zhuang}},\ }\href {\doibase
  10.1103/PhysRevC.89.054911} {\bibfield  {journal} {\bibinfo  {journal} {Phys.
  Rev. C}\ }\textbf {\bibinfo {volume} {89}},\ \bibinfo {pages} {054911}
  (\bibinfo {year} {2014})},\ \Eprint {http://arxiv.org/abs/1401.5845}
  {arXiv:1401.5845 [nucl-th]} \BibitemShut {NoStop}%
\bibitem [{\citenamefont {He}\ \emph {et~al.}(2022)\citenamefont {He},
  \citenamefont {Wu},\ and\ \citenamefont {Rapp}}]{He:2021zej}%
  \BibitemOpen
  \bibfield  {author} {\bibinfo {author} {\bibfnamefont {M.}~\bibnamefont
  {He}}, \bibinfo {author} {\bibfnamefont {B.}~\bibnamefont {Wu}}, \ and\
  \bibinfo {author} {\bibfnamefont {R.}~\bibnamefont {Rapp}},\ }\href {\doibase
  10.1103/PhysRevLett.128.162301} {\bibfield  {journal} {\bibinfo  {journal}
  {Phys. Rev. Lett.}\ }\textbf {\bibinfo {volume} {128}},\ \bibinfo {pages}
  {162301} (\bibinfo {year} {2022})},\ \Eprint
  {http://arxiv.org/abs/2111.13528} {arXiv:2111.13528 [nucl-th]} \BibitemShut
  {NoStop}%
\bibitem [{\citenamefont {Vogt}(2015)}]{Vogt:2015uba}%
  \BibitemOpen
  \bibfield  {author} {\bibinfo {author} {\bibfnamefont {R.}~\bibnamefont
  {Vogt}},\ }\href {\doibase 10.1103/PhysRevC.92.034909} {\bibfield  {journal}
  {\bibinfo  {journal} {Phys. Rev. C}\ }\textbf {\bibinfo {volume} {92}},\
  \bibinfo {pages} {034909} (\bibinfo {year} {2015})},\ \Eprint
  {http://arxiv.org/abs/1507.04418} {arXiv:1507.04418 [hep-ph]} \BibitemShut
  {NoStop}%
\bibitem [{\citenamefont {Zhang}\ \emph {et~al.}(2022)\citenamefont {Zhang},
  \citenamefont {Liao}, \citenamefont {Qin}, \citenamefont {Wang},\ and\
  \citenamefont {Xing}}]{Zhang:2022rby}%
  \BibitemOpen
  \bibfield  {author} {\bibinfo {author} {\bibfnamefont {S.-L.}\ \bibnamefont
  {Zhang}}, \bibinfo {author} {\bibfnamefont {J.}~\bibnamefont {Liao}},
  \bibinfo {author} {\bibfnamefont {G.-Y.}\ \bibnamefont {Qin}}, \bibinfo
  {author} {\bibfnamefont {E.}~\bibnamefont {Wang}}, \ and\ \bibinfo {author}
  {\bibfnamefont {H.}~\bibnamefont {Xing}},\ }\href@noop {} {\  (\bibinfo
  {year} {2022})},\ \Eprint {http://arxiv.org/abs/2208.08323} {arXiv:2208.08323
  [hep-ph]} \BibitemShut {NoStop}%
\bibitem [{\citenamefont {Li}\ \emph {et~al.}(2023)\citenamefont {Li},
  \citenamefont {Wang},\ and\ \citenamefont {Zhang}}]{Li:2022tcr}%
  \BibitemOpen
  \bibfield  {author} {\bibinfo {author} {\bibfnamefont {Y.}~\bibnamefont
  {Li}}, \bibinfo {author} {\bibfnamefont {S.}~\bibnamefont {Wang}}, \ and\
  \bibinfo {author} {\bibfnamefont {B.-W.}\ \bibnamefont {Zhang}},\ }\href
  {\doibase 10.1103/PhysRevC.108.024905} {\bibfield  {journal} {\bibinfo
  {journal} {Phys. Rev. C}\ }\textbf {\bibinfo {volume} {108}},\ \bibinfo
  {pages} {024905} (\bibinfo {year} {2023})},\ \Eprint
  {http://arxiv.org/abs/2209.00548} {arXiv:2209.00548 [hep-ph]} \BibitemShut
  {NoStop}%
\bibitem [{\citenamefont {He}\ \emph {et~al.}(2015)\citenamefont {He},
  \citenamefont {Luo}, \citenamefont {Wang},\ and\ \citenamefont
  {Zhu}}]{He:2015pra}%
  \BibitemOpen
  \bibfield  {author} {\bibinfo {author} {\bibfnamefont {Y.}~\bibnamefont
  {He}}, \bibinfo {author} {\bibfnamefont {T.}~\bibnamefont {Luo}}, \bibinfo
  {author} {\bibfnamefont {X.-N.}\ \bibnamefont {Wang}}, \ and\ \bibinfo
  {author} {\bibfnamefont {Y.}~\bibnamefont {Zhu}},\ }\href {\doibase
  10.1103/PhysRevC.91.054908} {\bibfield  {journal} {\bibinfo  {journal} {Phys.
  Rev. C}\ }\textbf {\bibinfo {volume} {91}},\ \bibinfo {pages} {054908}
  (\bibinfo {year} {2015})},\ \bibinfo {note} {[Erratum: Phys.Rev.C 97, 019902
  (2018)]},\ \Eprint {http://arxiv.org/abs/1503.03313} {arXiv:1503.03313
  [nucl-th]} \BibitemShut {NoStop}%
\bibitem [{\citenamefont {Cao}\ \emph {et~al.}(2016)\citenamefont {Cao},
  \citenamefont {Luo}, \citenamefont {Qin},\ and\ \citenamefont
  {Wang}}]{Cao:2016gvr}%
  \BibitemOpen
  \bibfield  {author} {\bibinfo {author} {\bibfnamefont {S.}~\bibnamefont
  {Cao}}, \bibinfo {author} {\bibfnamefont {T.}~\bibnamefont {Luo}}, \bibinfo
  {author} {\bibfnamefont {G.-Y.}\ \bibnamefont {Qin}}, \ and\ \bibinfo
  {author} {\bibfnamefont {X.-N.}\ \bibnamefont {Wang}},\ }\href {\doibase
  10.1103/PhysRevC.94.014909} {\bibfield  {journal} {\bibinfo  {journal} {Phys.
  Rev. C}\ }\textbf {\bibinfo {volume} {94}},\ \bibinfo {pages} {014909}
  (\bibinfo {year} {2016})},\ \Eprint {http://arxiv.org/abs/1605.06447}
  {arXiv:1605.06447 [nucl-th]} \BibitemShut {NoStop}%
\bibitem [{\citenamefont {Pang}\ \emph {et~al.}(2012)\citenamefont {Pang},
  \citenamefont {Wang},\ and\ \citenamefont {Wang}}]{Pang:2012he}%
  \BibitemOpen
  \bibfield  {author} {\bibinfo {author} {\bibfnamefont {L.}~\bibnamefont
  {Pang}}, \bibinfo {author} {\bibfnamefont {Q.}~\bibnamefont {Wang}}, \ and\
  \bibinfo {author} {\bibfnamefont {X.-N.}\ \bibnamefont {Wang}},\ }\href
  {\doibase 10.1103/PhysRevC.86.024911} {\bibfield  {journal} {\bibinfo
  {journal} {Phys. Rev. C}\ }\textbf {\bibinfo {volume} {86}},\ \bibinfo
  {pages} {024911} (\bibinfo {year} {2012})},\ \Eprint
  {http://arxiv.org/abs/1205.5019} {arXiv:1205.5019 [nucl-th]} \BibitemShut
  {NoStop}%
\bibitem [{\citenamefont {Collins}\ and\ \citenamefont
  {Soper}(1982)}]{Collins:1981uw}%
  \BibitemOpen
  \bibfield  {author} {\bibinfo {author} {\bibfnamefont {J.~C.}\ \bibnamefont
  {Collins}}\ and\ \bibinfo {author} {\bibfnamefont {D.~E.}\ \bibnamefont
  {Soper}},\ }\href {\doibase 10.1016/0550-3213(82)90021-9} {\bibfield
  {journal} {\bibinfo  {journal} {Nucl. Phys. B}\ }\textbf {\bibinfo {volume}
  {194}},\ \bibinfo {pages} {445} (\bibinfo {year} {1982})}\BibitemShut
  {NoStop}%
\bibitem [{\citenamefont {Kaufmann}\ \emph {et~al.}(2015)\citenamefont
  {Kaufmann}, \citenamefont {Mukherjee},\ and\ \citenamefont
  {Vogelsang}}]{Kaufmann:2015hma}%
  \BibitemOpen
  \bibfield  {author} {\bibinfo {author} {\bibfnamefont {T.}~\bibnamefont
  {Kaufmann}}, \bibinfo {author} {\bibfnamefont {A.}~\bibnamefont {Mukherjee}},
  \ and\ \bibinfo {author} {\bibfnamefont {W.}~\bibnamefont {Vogelsang}},\
  }\href {\doibase 10.1103/PhysRevD.92.054015} {\bibfield  {journal} {\bibinfo
  {journal} {Phys. Rev. D}\ }\textbf {\bibinfo {volume} {92}},\ \bibinfo
  {pages} {054015} (\bibinfo {year} {2015})},\ \bibinfo {note} {[Erratum:
  Phys.Rev.D 101, 079901 (2020)]},\ \Eprint {http://arxiv.org/abs/1506.01415}
  {arXiv:1506.01415 [hep-ph]} \BibitemShut {NoStop}%
\bibitem [{\citenamefont {Alwall}\ \emph {et~al.}(2014)\citenamefont {Alwall},
  \citenamefont {Frederix}, \citenamefont {Frixione}, \citenamefont {Hirschi},
  \citenamefont {Maltoni}, \citenamefont {Mattelaer}, \citenamefont {Shao},
  \citenamefont {Stelzer}, \citenamefont {Torrielli},\ and\ \citenamefont
  {Zaro}}]{Alwall:2014hca}%
  \BibitemOpen
  \bibfield  {author} {\bibinfo {author} {\bibfnamefont {J.}~\bibnamefont
  {Alwall}}, \bibinfo {author} {\bibfnamefont {R.}~\bibnamefont {Frederix}},
  \bibinfo {author} {\bibfnamefont {S.}~\bibnamefont {Frixione}}, \bibinfo
  {author} {\bibfnamefont {V.}~\bibnamefont {Hirschi}}, \bibinfo {author}
  {\bibfnamefont {F.}~\bibnamefont {Maltoni}}, \bibinfo {author} {\bibfnamefont
  {O.}~\bibnamefont {Mattelaer}}, \bibinfo {author} {\bibfnamefont {H.~S.}\
  \bibnamefont {Shao}}, \bibinfo {author} {\bibfnamefont {T.}~\bibnamefont
  {Stelzer}}, \bibinfo {author} {\bibfnamefont {P.}~\bibnamefont {Torrielli}},
  \ and\ \bibinfo {author} {\bibfnamefont {M.}~\bibnamefont {Zaro}},\ }\href
  {\doibase 10.1007/JHEP07(2014)079} {\bibfield  {journal} {\bibinfo  {journal}
  {JHEP}\ }\textbf {\bibinfo {volume} {07}},\ \bibinfo {pages} {079} (\bibinfo
  {year} {2014})},\ \Eprint {http://arxiv.org/abs/1405.0301} {arXiv:1405.0301
  [hep-ph]} \BibitemShut {NoStop}%
\bibitem [{\citenamefont {Dulat}\ \emph {et~al.}(2016)\citenamefont {Dulat},
  \citenamefont {Hou}, \citenamefont {Gao}, \citenamefont {Guzzi},
  \citenamefont {Huston}, \citenamefont {Nadolsky}, \citenamefont {Pumplin},
  \citenamefont {Schmidt}, \citenamefont {Stump},\ and\ \citenamefont
  {Yuan}}]{Dulat:2015mca}%
  \BibitemOpen
  \bibfield  {author} {\bibinfo {author} {\bibfnamefont {S.}~\bibnamefont
  {Dulat}}, \bibinfo {author} {\bibfnamefont {T.-J.}\ \bibnamefont {Hou}},
  \bibinfo {author} {\bibfnamefont {J.}~\bibnamefont {Gao}}, \bibinfo {author}
  {\bibfnamefont {M.}~\bibnamefont {Guzzi}}, \bibinfo {author} {\bibfnamefont
  {J.}~\bibnamefont {Huston}}, \bibinfo {author} {\bibfnamefont
  {P.}~\bibnamefont {Nadolsky}}, \bibinfo {author} {\bibfnamefont
  {J.}~\bibnamefont {Pumplin}}, \bibinfo {author} {\bibfnamefont
  {C.}~\bibnamefont {Schmidt}}, \bibinfo {author} {\bibfnamefont
  {D.}~\bibnamefont {Stump}}, \ and\ \bibinfo {author} {\bibfnamefont {C.~P.}\
  \bibnamefont {Yuan}},\ }\href {\doibase 10.1103/PhysRevD.93.033006}
  {\bibfield  {journal} {\bibinfo  {journal} {Phys. Rev. D}\ }\textbf {\bibinfo
  {volume} {93}},\ \bibinfo {pages} {033006} (\bibinfo {year} {2016})},\
  \Eprint {http://arxiv.org/abs/1506.07443} {arXiv:1506.07443 [hep-ph]}
  \BibitemShut {NoStop}%
\bibitem [{\citenamefont {Buckley}\ \emph {et~al.}(2015)\citenamefont
  {Buckley}, \citenamefont {Ferrando}, \citenamefont {Lloyd}, \citenamefont
  {Nordstr\"om}, \citenamefont {Page}, \citenamefont {R\"ufenacht},
  \citenamefont {Sch\"onherr},\ and\ \citenamefont {Watt}}]{Buckley:2014ana}%
  \BibitemOpen
  \bibfield  {author} {\bibinfo {author} {\bibfnamefont {A.}~\bibnamefont
  {Buckley}}, \bibinfo {author} {\bibfnamefont {J.}~\bibnamefont {Ferrando}},
  \bibinfo {author} {\bibfnamefont {S.}~\bibnamefont {Lloyd}}, \bibinfo
  {author} {\bibfnamefont {K.}~\bibnamefont {Nordstr\"om}}, \bibinfo {author}
  {\bibfnamefont {B.}~\bibnamefont {Page}}, \bibinfo {author} {\bibfnamefont
  {M.}~\bibnamefont {R\"ufenacht}}, \bibinfo {author} {\bibfnamefont
  {M.}~\bibnamefont {Sch\"onherr}}, \ and\ \bibinfo {author} {\bibfnamefont
  {G.}~\bibnamefont {Watt}},\ }\href {\doibase 10.1140/epjc/s10052-015-3318-8}
  {\bibfield  {journal} {\bibinfo  {journal} {Eur. Phys. J. C}\ }\textbf
  {\bibinfo {volume} {75}},\ \bibinfo {pages} {132} (\bibinfo {year} {2015})},\
  \Eprint {http://arxiv.org/abs/1412.7420} {arXiv:1412.7420 [hep-ph]}
  \BibitemShut {NoStop}%
\bibitem [{\citenamefont {Ma}\ \emph {et~al.}(2014)\citenamefont {Ma},
  \citenamefont {Qiu},\ and\ \citenamefont {Zhang}}]{Ma:2013yla}%
  \BibitemOpen
  \bibfield  {author} {\bibinfo {author} {\bibfnamefont {Y.-Q.}\ \bibnamefont
  {Ma}}, \bibinfo {author} {\bibfnamefont {J.-W.}\ \bibnamefont {Qiu}}, \ and\
  \bibinfo {author} {\bibfnamefont {H.}~\bibnamefont {Zhang}},\ }\href
  {\doibase 10.1103/PhysRevD.89.094029} {\bibfield  {journal} {\bibinfo
  {journal} {Phys. Rev. D}\ }\textbf {\bibinfo {volume} {89}},\ \bibinfo
  {pages} {094029} (\bibinfo {year} {2014})},\ \Eprint
  {http://arxiv.org/abs/1311.7078} {arXiv:1311.7078 [hep-ph]} \BibitemShut
  {NoStop}%
\bibitem [{\citenamefont {Bodwin}\ and\ \citenamefont
  {Lee}(2004)}]{Bodwin:2003wh}%
  \BibitemOpen
  \bibfield  {author} {\bibinfo {author} {\bibfnamefont {G.~T.}\ \bibnamefont
  {Bodwin}}\ and\ \bibinfo {author} {\bibfnamefont {J.}~\bibnamefont {Lee}},\
  }\href {\doibase 10.1103/PhysRevD.69.054003} {\bibfield  {journal} {\bibinfo
  {journal} {Phys. Rev. D}\ }\textbf {\bibinfo {volume} {69}},\ \bibinfo
  {pages} {054003} (\bibinfo {year} {2004})},\ \Eprint
  {http://arxiv.org/abs/hep-ph/0308016} {arXiv:hep-ph/0308016} \BibitemShut
  {NoStop}%
\bibitem [{\citenamefont {Bodwin}\ \emph {et~al.}(2014)\citenamefont {Bodwin},
  \citenamefont {Chung}, \citenamefont {Kim},\ and\ \citenamefont
  {Lee}}]{Bodwin:2014gia}%
  \BibitemOpen
  \bibfield  {author} {\bibinfo {author} {\bibfnamefont {G.~T.}\ \bibnamefont
  {Bodwin}}, \bibinfo {author} {\bibfnamefont {H.~S.}\ \bibnamefont {Chung}},
  \bibinfo {author} {\bibfnamefont {U.-R.}\ \bibnamefont {Kim}}, \ and\
  \bibinfo {author} {\bibfnamefont {J.}~\bibnamefont {Lee}},\ }\href {\doibase
  10.1103/PhysRevLett.113.022001} {\bibfield  {journal} {\bibinfo  {journal}
  {Phys. Rev. Lett.}\ }\textbf {\bibinfo {volume} {113}},\ \bibinfo {pages}
  {022001} (\bibinfo {year} {2014})},\ \Eprint {http://arxiv.org/abs/1403.3612}
  {arXiv:1403.3612 [hep-ph]} \BibitemShut {NoStop}%
\bibitem [{\citenamefont {Bodwin}\ \emph {et~al.}(2016)\citenamefont {Bodwin},
  \citenamefont {Chao}, \citenamefont {Chung}, \citenamefont {Kim},
  \citenamefont {Lee},\ and\ \citenamefont {Ma}}]{Bodwin:2015iua}%
  \BibitemOpen
  \bibfield  {author} {\bibinfo {author} {\bibfnamefont {G.~T.}\ \bibnamefont
  {Bodwin}}, \bibinfo {author} {\bibfnamefont {K.-T.}\ \bibnamefont {Chao}},
  \bibinfo {author} {\bibfnamefont {H.~S.}\ \bibnamefont {Chung}}, \bibinfo
  {author} {\bibfnamefont {U.-R.}\ \bibnamefont {Kim}}, \bibinfo {author}
  {\bibfnamefont {J.}~\bibnamefont {Lee}}, \ and\ \bibinfo {author}
  {\bibfnamefont {Y.-Q.}\ \bibnamefont {Ma}},\ }\href {\doibase
  10.1103/PhysRevD.93.034041} {\bibfield  {journal} {\bibinfo  {journal} {Phys.
  Rev. D}\ }\textbf {\bibinfo {volume} {93}},\ \bibinfo {pages} {034041}
  (\bibinfo {year} {2016})},\ \Eprint {http://arxiv.org/abs/1509.07904}
  {arXiv:1509.07904 [hep-ph]} \BibitemShut {NoStop}%
\bibitem [{\citenamefont {Gong}\ \emph {et~al.}(2013)\citenamefont {Gong},
  \citenamefont {Wan}, \citenamefont {Wang},\ and\ \citenamefont
  {Zhang}}]{Gong:2012ug}%
  \BibitemOpen
  \bibfield  {author} {\bibinfo {author} {\bibfnamefont {B.}~\bibnamefont
  {Gong}}, \bibinfo {author} {\bibfnamefont {L.-P.}\ \bibnamefont {Wan}},
  \bibinfo {author} {\bibfnamefont {J.-X.}\ \bibnamefont {Wang}}, \ and\
  \bibinfo {author} {\bibfnamefont {H.-F.}\ \bibnamefont {Zhang}},\ }\href
  {\doibase 10.1103/PhysRevLett.110.042002} {\bibfield  {journal} {\bibinfo
  {journal} {Phys. Rev. Lett.}\ }\textbf {\bibinfo {volume} {110}},\ \bibinfo
  {pages} {042002} (\bibinfo {year} {2013})},\ \Eprint
  {http://arxiv.org/abs/1205.6682} {arXiv:1205.6682 [hep-ph]} \BibitemShut
  {NoStop}%
\bibitem [{\citenamefont {Chao}\ \emph {et~al.}(2012)\citenamefont {Chao},
  \citenamefont {Ma}, \citenamefont {Shao}, \citenamefont {Wang},\ and\
  \citenamefont {Zhang}}]{Chao:2012iv}%
  \BibitemOpen
  \bibfield  {author} {\bibinfo {author} {\bibfnamefont {K.-T.}\ \bibnamefont
  {Chao}}, \bibinfo {author} {\bibfnamefont {Y.-Q.}\ \bibnamefont {Ma}},
  \bibinfo {author} {\bibfnamefont {H.-S.}\ \bibnamefont {Shao}}, \bibinfo
  {author} {\bibfnamefont {K.}~\bibnamefont {Wang}}, \ and\ \bibinfo {author}
  {\bibfnamefont {Y.-J.}\ \bibnamefont {Zhang}},\ }\href {\doibase
  10.1103/PhysRevLett.108.242004} {\bibfield  {journal} {\bibinfo  {journal}
  {Phys. Rev. Lett.}\ }\textbf {\bibinfo {volume} {108}},\ \bibinfo {pages}
  {242004} (\bibinfo {year} {2012})},\ \Eprint {http://arxiv.org/abs/1201.2675}
  {arXiv:1201.2675 [hep-ph]} \BibitemShut {NoStop}%
\bibitem [{\citenamefont {Ma}\ \emph {et~al.}(2011)\citenamefont {Ma},
  \citenamefont {Wang},\ and\ \citenamefont {Chao}}]{Ma:2010jj}%
  \BibitemOpen
  \bibfield  {author} {\bibinfo {author} {\bibfnamefont {Y.-Q.}\ \bibnamefont
  {Ma}}, \bibinfo {author} {\bibfnamefont {K.}~\bibnamefont {Wang}}, \ and\
  \bibinfo {author} {\bibfnamefont {K.-T.}\ \bibnamefont {Chao}},\ }\href
  {\doibase 10.1103/PhysRevD.84.114001} {\bibfield  {journal} {\bibinfo
  {journal} {Phys. Rev. D}\ }\textbf {\bibinfo {volume} {84}},\ \bibinfo
  {pages} {114001} (\bibinfo {year} {2011})},\ \Eprint
  {http://arxiv.org/abs/1012.1030} {arXiv:1012.1030 [hep-ph]} \BibitemShut
  {NoStop}%
\bibitem [{\citenamefont {Brambilla}\ \emph {et~al.}(2023)\citenamefont
  {Brambilla}, \citenamefont {Chung}, \citenamefont {Vairo},\ and\
  \citenamefont {Wang}}]{Brambilla:2022ayc}%
  \BibitemOpen
  \bibfield  {author} {\bibinfo {author} {\bibfnamefont {N.}~\bibnamefont
  {Brambilla}}, \bibinfo {author} {\bibfnamefont {H.~S.}\ \bibnamefont
  {Chung}}, \bibinfo {author} {\bibfnamefont {A.}~\bibnamefont {Vairo}}, \ and\
  \bibinfo {author} {\bibfnamefont {X.-P.}\ \bibnamefont {Wang}},\ }\href
  {\doibase 10.1007/JHEP03(2023)242} {\bibfield  {journal} {\bibinfo  {journal}
  {JHEP}\ }\textbf {\bibinfo {volume} {03}},\ \bibinfo {pages} {242} (\bibinfo
  {year} {2023})},\ \Eprint {http://arxiv.org/abs/2210.17345} {arXiv:2210.17345
  [hep-ph]} \BibitemShut {NoStop}%
\bibitem [{\citenamefont {Gao}\ \emph {et~al.}(2024)\citenamefont {Gao},
  \citenamefont {Liu}, \citenamefont {Shen}, \citenamefont {Xing},\ and\
  \citenamefont {Zhao}}]{Gao:2024nkz}%
  \BibitemOpen
  \bibfield  {author} {\bibinfo {author} {\bibfnamefont {J.}~\bibnamefont
  {Gao}}, \bibinfo {author} {\bibfnamefont {C.}~\bibnamefont {Liu}}, \bibinfo
  {author} {\bibfnamefont {X.}~\bibnamefont {Shen}}, \bibinfo {author}
  {\bibfnamefont {H.}~\bibnamefont {Xing}}, \ and\ \bibinfo {author}
  {\bibfnamefont {Y.}~\bibnamefont {Zhao}},\ }\href@noop {} {\  (\bibinfo
  {year} {2024})},\ \Eprint {http://arxiv.org/abs/2401.02781} {arXiv:2401.02781
  [hep-ph]} \BibitemShut {NoStop}%
\bibitem [{\citenamefont {Anderle}\ \emph {et~al.}(2017)\citenamefont
  {Anderle}, \citenamefont {Kaufmann}, \citenamefont {Stratmann}, \citenamefont
  {Ringer},\ and\ \citenamefont {Vitev}}]{Anderle:2017cgl}%
  \BibitemOpen
  \bibfield  {author} {\bibinfo {author} {\bibfnamefont {D.~P.}\ \bibnamefont
  {Anderle}}, \bibinfo {author} {\bibfnamefont {T.}~\bibnamefont {Kaufmann}},
  \bibinfo {author} {\bibfnamefont {M.}~\bibnamefont {Stratmann}}, \bibinfo
  {author} {\bibfnamefont {F.}~\bibnamefont {Ringer}}, \ and\ \bibinfo {author}
  {\bibfnamefont {I.}~\bibnamefont {Vitev}},\ }\href {\doibase
  10.1103/PhysRevD.96.034028} {\bibfield  {journal} {\bibinfo  {journal} {Phys.
  Rev. D}\ }\textbf {\bibinfo {volume} {96}},\ \bibinfo {pages} {034028}
  (\bibinfo {year} {2017})},\ \Eprint {http://arxiv.org/abs/1706.09857}
  {arXiv:1706.09857 [hep-ph]} \BibitemShut {NoStop}%
\bibitem [{\citenamefont {Guo}\ and\ \citenamefont {Wang}(2000)}]{Guo:2000nz}%
  \BibitemOpen
  \bibfield  {author} {\bibinfo {author} {\bibfnamefont {X.-f.}\ \bibnamefont
  {Guo}}\ and\ \bibinfo {author} {\bibfnamefont {X.-N.}\ \bibnamefont {Wang}},\
  }\href {\doibase 10.1103/PhysRevLett.85.3591} {\bibfield  {journal} {\bibinfo
   {journal} {Phys. Rev. Lett.}\ }\textbf {\bibinfo {volume} {85}},\ \bibinfo
  {pages} {3591} (\bibinfo {year} {2000})},\ \Eprint
  {http://arxiv.org/abs/hep-ph/0005044} {arXiv:hep-ph/0005044} \BibitemShut
  {NoStop}%
\bibitem [{\citenamefont {Zhang}\ \emph {et~al.}(2004)\citenamefont {Zhang},
  \citenamefont {Wang},\ and\ \citenamefont {Wang}}]{Zhang:2003wk}%
  \BibitemOpen
  \bibfield  {author} {\bibinfo {author} {\bibfnamefont {B.-W.}\ \bibnamefont
  {Zhang}}, \bibinfo {author} {\bibfnamefont {E.}~\bibnamefont {Wang}}, \ and\
  \bibinfo {author} {\bibfnamefont {X.-N.}\ \bibnamefont {Wang}},\ }\href
  {\doibase 10.1103/PhysRevLett.93.072301} {\bibfield  {journal} {\bibinfo
  {journal} {Phys. Rev. Lett.}\ }\textbf {\bibinfo {volume} {93}},\ \bibinfo
  {pages} {072301} (\bibinfo {year} {2004})},\ \Eprint
  {http://arxiv.org/abs/nucl-th/0309040} {arXiv:nucl-th/0309040} \BibitemShut
  {NoStop}%
\bibitem [{\citenamefont {Luo}\ \emph {et~al.}(2023)\citenamefont {Luo},
  \citenamefont {He}, \citenamefont {Cao},\ and\ \citenamefont
  {Wang}}]{Luo:2023nsi}%
  \BibitemOpen
  \bibfield  {author} {\bibinfo {author} {\bibfnamefont {T.}~\bibnamefont
  {Luo}}, \bibinfo {author} {\bibfnamefont {Y.}~\bibnamefont {He}}, \bibinfo
  {author} {\bibfnamefont {S.}~\bibnamefont {Cao}}, \ and\ \bibinfo {author}
  {\bibfnamefont {X.-N.}\ \bibnamefont {Wang}},\ }\href@noop {} {\  (\bibinfo
  {year} {2023})},\ \Eprint {http://arxiv.org/abs/2306.13742} {arXiv:2306.13742
  [nucl-th]} \BibitemShut {NoStop}%
\bibitem [{\citenamefont {Zhang}\ \emph {et~al.}(2018)\citenamefont {Zhang},
  \citenamefont {Luo}, \citenamefont {Wang},\ and\ \citenamefont
  {Zhang}}]{Zhang:2018urd}%
  \BibitemOpen
  \bibfield  {author} {\bibinfo {author} {\bibfnamefont {S.-L.}\ \bibnamefont
  {Zhang}}, \bibinfo {author} {\bibfnamefont {T.}~\bibnamefont {Luo}}, \bibinfo
  {author} {\bibfnamefont {X.-N.}\ \bibnamefont {Wang}}, \ and\ \bibinfo
  {author} {\bibfnamefont {B.-W.}\ \bibnamefont {Zhang}},\ }\href {\doibase
  10.1103/PhysRevC.98.021901} {\bibfield  {journal} {\bibinfo  {journal} {Phys.
  Rev. C}\ }\textbf {\bibinfo {volume} {98}},\ \bibinfo {pages} {021901}
  (\bibinfo {year} {2018})},\ \Eprint {http://arxiv.org/abs/1804.11041}
  {arXiv:1804.11041 [nucl-th]} \BibitemShut {NoStop}%
\bibitem [{\citenamefont {Caucal}\ \emph {et~al.}(2018)\citenamefont {Caucal},
  \citenamefont {Iancu}, \citenamefont {Mueller},\ and\ \citenamefont
  {Soyez}}]{Caucal:2018dla}%
  \BibitemOpen
  \bibfield  {author} {\bibinfo {author} {\bibfnamefont {P.}~\bibnamefont
  {Caucal}}, \bibinfo {author} {\bibfnamefont {E.}~\bibnamefont {Iancu}},
  \bibinfo {author} {\bibfnamefont {A.~H.}\ \bibnamefont {Mueller}}, \ and\
  \bibinfo {author} {\bibfnamefont {G.}~\bibnamefont {Soyez}},\ }\href
  {\doibase 10.1103/PhysRevLett.120.232001} {\bibfield  {journal} {\bibinfo
  {journal} {Phys. Rev. Lett.}\ }\textbf {\bibinfo {volume} {120}},\ \bibinfo
  {pages} {232001} (\bibinfo {year} {2018})},\ \Eprint
  {http://arxiv.org/abs/1801.09703} {arXiv:1801.09703 [hep-ph]} \BibitemShut
  {NoStop}%
\bibitem [{\citenamefont {Xing}\ \emph {et~al.}(2024)\citenamefont {Xing},
  \citenamefont {Cao},\ and\ \citenamefont {Qin}}]{Xing:2023ciw}%
  \BibitemOpen
  \bibfield  {author} {\bibinfo {author} {\bibfnamefont {W.-J.}\ \bibnamefont
  {Xing}}, \bibinfo {author} {\bibfnamefont {S.}~\bibnamefont {Cao}}, \ and\
  \bibinfo {author} {\bibfnamefont {G.-Y.}\ \bibnamefont {Qin}},\ }\href
  {\doibase 10.1016/j.physletb.2024.138523} {\bibfield  {journal} {\bibinfo
  {journal} {Phys. Lett. B}\ }\textbf {\bibinfo {volume} {850}},\ \bibinfo
  {pages} {138523} (\bibinfo {year} {2024})},\ \Eprint
  {http://arxiv.org/abs/2303.12485} {arXiv:2303.12485 [hep-ph]} \BibitemShut
  {NoStop}%
\bibitem [{\citenamefont {Pablos}(2020)}]{Pablos:2019ngg}%
  \BibitemOpen
  \bibfield  {author} {\bibinfo {author} {\bibfnamefont {D.}~\bibnamefont
  {Pablos}},\ }\href {\doibase 10.1103/PhysRevLett.124.052301} {\bibfield
  {journal} {\bibinfo  {journal} {Phys. Rev. Lett.}\ }\textbf {\bibinfo
  {volume} {124}},\ \bibinfo {pages} {052301} (\bibinfo {year} {2020})},\
  \Eprint {http://arxiv.org/abs/1907.12301} {arXiv:1907.12301 [hep-ph]}
  \BibitemShut {NoStop}%
\bibitem [{\citenamefont {Zhang}\ \emph {et~al.}(2024)\citenamefont {Zhang},
  \citenamefont {Wang}, \citenamefont {Xing},\ and\ \citenamefont
  {Zhang}}]{Zhang:2023oid}%
  \BibitemOpen
  \bibfield  {author} {\bibinfo {author} {\bibfnamefont {S.-L.}\ \bibnamefont
  {Zhang}}, \bibinfo {author} {\bibfnamefont {E.}~\bibnamefont {Wang}},
  \bibinfo {author} {\bibfnamefont {H.}~\bibnamefont {Xing}}, \ and\ \bibinfo
  {author} {\bibfnamefont {B.-W.}\ \bibnamefont {Zhang}},\ }\href {\doibase
  10.1016/j.physletb.2024.138549} {\bibfield  {journal} {\bibinfo  {journal}
  {Phys. Lett. B}\ }\textbf {\bibinfo {volume} {850}},\ \bibinfo {pages}
  {138549} (\bibinfo {year} {2024})},\ \Eprint
  {http://arxiv.org/abs/2303.14881} {arXiv:2303.14881 [hep-ph]} \BibitemShut
  {NoStop}%
\end{thebibliography}%

\end{document}